\documentclass[10pt,aps,prc,twocolumn,amsmath,superscriptaddress,floatfix,nofootinbib]{revtex4-2}
\usepackage{setspace,ulem,amssymb,amsfonts,amsmath,mathtools,bm,color,xcolor,graphicx,braket,adjustbox,esint,upgreek,verbatim,ctable,placeins}

\usepackage[colorlinks, linkcolor=red,anchorcolor=green,citecolor=blue]{hyperref}

\usepackage[colorlinks, linkcolor=red,anchorcolor=green,citecolor=blue]{hyperref}

\usepackage{hyperref}
\hypersetup{colorlinks=true, urlcolor=blue}

\begin{document}

\title{Data-driven analysis for the bottomonium potential in the quark-gluon plasma}

\author{Shuhan Zheng}
\email{Contact author: zhengsh24@mails.tsinghua.edu.cn}
\affiliation{Department of Physics, Tsinghua University, Beijing 100084, China}
\affiliation{Department of Physics, Tianjin University, Tianjin 300354, China}

\author{Baoyi Chen}
\email{Contact author: baoyi.chen@tju.edu.cn}
\affiliation{Department of Physics, Tianjin University, Tianjin 300354, China}

\author{Xiaojian Du}
\email{Contact author: xiaojiandu@outlook.com}
\affiliation{Wilczek Quantum Center, Shanghai Institute for Advanced Studies,\\
University of Science and Technology of China, Shanghai 201315, China}
\affiliation{Shanghai Research Center for Quantum Sciences, Shanghai 201315, China}
\affiliation{Hefei National Laboratory, Hefei 230088, China}
\affiliation{Department of Physics, University of Jyväskylä, P. O. Box 35, 40014 Jyväskylä, Finland}
\affiliation{Helsinki Institute of Physics, University of Helsinki, P. O. Box 64, 00014 Helsinki, Finland}

\author{Shuzhe Shi}
\email{Contact author: shuzhe-shi@tsinghua.edu.cn}
\affiliation{Department of Physics, Tsinghua University, Beijing 100084, China}
\affiliation{State Key Laboratory of Low-Dimensional Quantum Physics, Tsinghua University, Beijing 100084, China}

\begin{abstract}
We present a data-driven analysis within a quantum evolutionary microscopic framework to constrain the in-medium bottomonium potential. In relativistic heavy-ion collisions, bottomonium bound states serve as invaluable probes of the quark-gluon plasma (QGP) owing to their negligible production in the QGP phase. Meanwhile, their nonrelativistic nature allows a straightforward theoretical description via effective field theories such as potential models. Recent lattice QCD calculations of the bottomonium interaction potential have yielded qualitatively distinct results. These discrepancies motivate a data-driven extraction of the potential based on heavy-ion experiments. In this work, we perform a Bayesian analysis to constrain the bottomonium interaction potential. The relationship between potential parameters and observables is established by numerically solving the nonrelativistic time-dependent Schr\"odinger equation. By comparing these simulations with experimental measurements, our Bayesian framework provides the effective potential that is readily testable in future experiments.
\end{abstract}
\date{\today}
\maketitle
\section{Introduction}
The study of quark-gluon plasma (QGP) is a central topic in high-energy nuclear physics. Among the probes of the QGP, quarkonium bound states -- mesons with the constituent quark being a charm or bottom quark and its corresponding antiquark---play a crucial role because of the clear interpretation of in-medium modification and  production, benefiting from their large mass~\cite{Matsui:1986dk, Karsch:1987pv, Brambilla:2010cs, Mocsy:2013syh, Grandchamp:2003uw, Song:2011xi, Emerick:2011xu, Du:2017qkv}. Particularly, the nonrelativistic nature of quarkonium allows for a theoretical description using effective field theories, such as the potential nonrelativistic quantum chromodynamics (pNRQCD)~\cite{Brambilla:1999xf, Brambilla:2008cx, Brambilla:2010vq}, where the in-medium interaction between heavy quarks is described by a temperature-dependent, complex-valued, static potential. The real part of the potential receives a distance-dependent reduction compared to the vacuum values, caused by the medium's screening effect; a nonzero imaginary potential also emerges because of the in-medium transition from color-singlet to color-octet states.
Such qualitative properties are seen in pNRQCD calculations~\cite{Brambilla:1999xf, Brambilla:2008cx, Brambilla:2010vq}, one-loop hard thermal loop perturbative QCD calculations~\cite{Laine:2006ns, Beraudo:2007ky}, and lattice QCD calculations~\cite{Rothkopf:2011db, Burnier:2014ssa, Burnier:2015tda, Bala:2019cqu, Bala:2021fkm, Bazavov:2023dci, Ali:2025iux}.
Thus, one may solve the time-dependent Schr\"odinger equation with the in-medium interaction potential to model the quarkonium production in relativistic heavy-ion collisions~\cite{Satz:2005hx, Liu:2010ej, Chen:2016vha, Wen:2022utn, Krouppa:2015yoa, Krouppa:2017jlg, Strickland:2023nfm}.

Although lattice QCD simulations provide nonperturbative calculations of the heavy-quark potentials, the numerical results rely on specific model assumptions in the extraction, owing to the ill-posed nature of reconstructing (real time) potential from imaginary-time correlators~\cite{Shi:2022yqw}. Indeed, lattice QCD calculations of bottomonium potentials fall into two groups, one expects strong screening in the real part and modest magnitude in the imaginary potential~\cite{Rothkopf:2011db, Burnier:2014ssa, Burnier:2015tda}, the other exhibits mild screening effect but strong enhancement of the imaginary part~\cite{Bala:2019cqu, Bala:2021fkm, Bazavov:2023dci, Ali:2025iux}.
In this work, our aim is to address such a discrepancy by a data-drive extraction of the interaction potential. A Bayesian analysis of the real and imaginary potentials is performed on the basis of the experimental measurements of the nuclear modification factors of ground and excited bottomonium states. 

Bayesian analysis, widely used to extract QCD properties in heavy-ion collisions~\cite{Bernhard:2016tnd, Bernhard:2019bmu, JETSCAPE:2020shq, JETSCAPE:2020mzn, Heffernan:2023gye, Heffernan:2023utr, Xie:2022ght, Xie:2022fak, Gonzalez:2020bqm, MUSES:2023hyz, Domingues:2024pom, Grishmanovskii:2025mnc, Altenkort:2023eav, Liu:2023rfi, Jahan:2024wpj, Nijs:2020ors, Nijs:2020roc, Devetak:2019lsk, Guo:2025tsf, JETSCAPE:2021ehl, JETSCAPE:2024cqe, Andronic:2025ylc} and other fields, provides a statistical description of model parameters based on the level of agreement with the experimental data and prior knowledge of the model. In the realm of heavy-quark physics, a data-driven Langevin transport framework combined with Bayesian inference was employed to constrain temperature-dependent diffusion coefficients, showing that simultaneous fits to $D$-meson $R_{AA}$ and $v_2$ data can robustly determine the heavy-quark potential parameters~\cite{Li:2019lex}.

The structure of this paper is as follows. The methodology of simulation of bottomonium production in heavy-ion collisions will be thoroughly demonstrated in Sec.~\ref{sec:schroedinger}, followed by the details for the Bayesian analysis in Sec.~\ref{sec:bayesian}. Section~\ref{sec:results} provides the results of the data-driven analysis by displaying the posterior likelihood distribution and the simulation results of the nuclear modification factor, with the corresponding parameterized potential also being discussed.
After the summary and outlook in Sec.~\ref{sec:summary}, the relevant Appendixes will finally be given, including details on numerically solving the time-dependent equation Schr\"odinger (Appendix~\ref{sec:app:tma}), the model emulator (Appendix~\ref{sec:app:emulator}) and the sampling of potential parameters (Appendix~\ref{sec:app:mcmc}). In this paper, each equation will be expressed with the natural unit system, where $k = c= \hbar=1$.

\section{Simulation of Bottomonium Evolutions with Nonrelativistic Complex Potential Model} \label{sec:schroedinger}
In this study, the effects of the hot medium are included in the temperature-dependent complex potential in the evolution of bottomonium~\cite{Guo:2012hx}. Due to the large mass of the bottom quarks compared to the kinetic energy of the inner motion of bottomonium, one can neglect the relativistic effect and describe the evolution of bottomonium with the Schr\"odinger equation. We follow the procedure in, e.g., Refs.~\cite{Islam:2020gdv, Wen:2022yjx}. Neglecting the effect of medium viscosity on the heavy-quark potential, both the real and imaginary parts of the potential become isotropic. The interaction potential ($V$) depends only on the radial distance ($r$) and temperature ($T$), not on the azimuthal angles $(\theta, \varphi)$. In this case, there are no mixing between bottomonium eigenstates with different angular quantum numbers. Both the wave functions and the Schr\"odinger equation can be separated into radial and angular components, the latter of which explicitly depends on the angular ($\ell$) and magnetic ($m$) quantum numbers, and the total wave function reads:
\begin{align}
    \psi_{\ell,m}(\boldsymbol{r},t) = R_\ell(r,t) Y_{\ell,m}(\theta,\varphi),
\end{align}
in which the radial part can be expressed by the following formula:
\begin{align}
&i \frac{\partial}{\partial t}u_\ell(r, t) = \notag \\
&\left[-\frac{1}{2 m_{\mu}} \frac{\partial^{2}}{\partial r^{2}}+V(r,T(t))+\frac{\ell(\ell+1)}{2 m_{\mu} r^{2}}\right] u_\ell(r, t), \label{def_Sch_Eq}
\end{align}
where $r$ represents the relative distance between the bottom and antibottom quarks and $u_\ell(r,t)= r R_\ell(r,t)$ is the product of $r$ and the radial component of the wave function $R_\ell(r,t)$.
The mass of a single bottom quark $m_b$ is selected as 4.62 GeV, and $m_{\mu} = m_b/2$ is the reduced mass. To describe the dissociation of the bottomonium bound state~\cite{Brambilla:2024tqg} and the state transition from the singlet to the color octets, the form of the effective potential can then be written in the following equation:
\begin{equation}
V(r, T(t)) = V_R(r,T(t))  - i\,V_I(r,T(t)).
\end{equation}
In our simulation, $T(t)$ in the temperature-dependent interaction potential $V(r, T(t))$ is the temperature that the $b\bar{b}$ state feels at its current position $\mathbf{x}(t)=\mathbf{x}_0 + \mathbf{v}\,t$, i.e., $T(t) = T(t, \mathbf{x}(t))$. $\mathbf{x}_0$ is the initial position of the $b\bar{b}$ state and $\mathbf{v}$ represents its velocity, which is assumed to be constant during evolution. The space-time profile of the temperature $T(t,\mathbf{x})$ is given by the hydrodynamic background, which can be obtained by hydrodynamic simulation using the \textsc{music} package~\cite{Schenke:2010rr, Schenke:2010nt} with taking the \textsc{s95p} equation of state~\cite{Huovinen:2009yb}, a constant shear viscosity $\eta/s = 0.08$, and vanishing bulk viscosity~\cite{Bernhard:2016tnd}, as well as the event-averaged Glauber initial condition~\cite{Miller:2007ri} tuned to reproduce light-hadron production. It might be worth noting that the evolution of heavy flavor particles is insensitive to the details of the hydrodynamic background, which is reflected in Ref.~\cite{Rapp:2018qla}.
In our computation, we focus on mid-rapidity particles and have taken the longitudinal boost-invariant hydrodynamic background. Thus, we focus on bottomonium particles with vanishing rapidity, $y = 0$.
Meanwhile, the backreaction to the hydrodynamic profile from bottomonium is neglected.

We start the in-medium evolution of the bottomonium at $t_{\mathrm{init}} = 1~\mathrm{fm}$, determined by the formation times of the $b\bar{b}$ state and the QGP fluid.
We found that varying $t_{\mathrm{init}}$ by $\sim 0.4~\mathrm{fm}$ results in final yields of bottomonium eigenstates that lie within the uncertainty arising from other sources.
The interval evolution of $b\bar b$ dipoles ceases when the local temperature of QGP falls below a critical value, referred to as the switching temperature $T_d$ in this work. The switching temperature is defined as the value at which the in-medium bottomonium potential reverts to the vacuum Cornell potential, meaning the imaginary part vanishes and the real part coincides with the Cornell potential simultaneously. We denote the switching time by $t_d$, which, as a consequence, depends on the initial position and velocity of the bottomonium.
Although $T_d$ can be chosen naturally as the pseudocritical temperature, $160~\mathrm{MeV}$, of the transition between QGP and hadron gas, the crossover nature of the phase transition allows for a different value.
Thus, in this work, we adopt three choices $T_d = \{160,180,200 \}~\mathrm{MeV}$.

After switching, the probability of obtaining a bottomonium state, i.e., the vacuum eigenstate with principal quantum number $n$ and orbital angular number $\ell$, is given by
\begin{align}
    P_{n,\ell} = |c_{n,\ell}|^2\,,
    \label{eq:probability_nl}
\end{align}
where $c_{n,\ell} = \int_0^{\infty} r^2 \mathrm{d}r~ \Psi_{n,\ell}^*(r) R_\ell (r,t_d)$ is the overlap coefficient between the final wave function $R_\ell (r,t_d)$ and the specific quantum state $\Psi_{n,\ell}$ in vacuum.

\subsection{Calculation of experimental observables}
With a complex potential, we can compute the survival probability by solving the time-dependent Schr\"odinger equation~\eqref{def_Sch_Eq}. This enables us to calculate the experimental observables, i.e., the nuclear modification factor ($R_{AA}$) and the elliptic flow ($v_2$), depending on the centrality of the collisions, the transverse momentum ($p_T$), and the rapidity interval ($y$). 

The angular-dependent nuclear modification factor is defined as the ratio between the differential production rate of the bottomonium bound state $(n,l)$ in AA collisions ($d^2N_{n\ell}^{AA}/dp_T d\varphi$) and that in pp collisions ($d^2N_{n\ell}^{pp}/dp_T d\varphi$) scaled by the number of binary collisions ($N_{\mathrm{coll}}$),
\begin{align}
R_{AA}^{n\ell}(p_T,\varphi) = \frac{d^2N_{n\ell}^{AA}/dp_T d\varphi}{N_{\mathrm{coll}}\,d^2N_{n\ell}^{pp}/dp_T d\varphi}.
\end{align}

It naturally derives the angular-averaged nuclear modification factor and the elliptic flow,
\begin{align}
    R_{AA}^{n\ell}(p_T) \equiv\,& \int R_{AA}^{n\ell}(p_T,\varphi) \frac{d\varphi}{2\pi}\,,\\
    v_{2}^{n\ell}(p_T) \equiv\,& \frac{\int R_{AA}^{n\ell}(p_T,\varphi) \cos(2\varphi)\frac{d\varphi}{2\pi}}{R_{AA}^{n\ell}(p_T)}\,.
\end{align}
As in the experiments, we are interested in the ``prompt'' particle production which includes the primordial production of these bound states and the feed-down from the short-lived higher excitations of bottomonium.
The production of the final-state ``prompt'' particles is formulated as~\cite{Strickland:2023nfm}
\begin{align}
&\frac{d^2N^{AA}_i}{dp_Td\varphi} = F_{i,j}\, \frac{d^2N^{\mathrm{prim}}_{AA,j}}{dp_Td\varphi}, \quad  
\frac{d^2N^{pp}_i}{dp_Td\varphi} = F_{i,j}\, \frac{d^2N^{\mathrm{prim}}_{pp,j}}{dp_Td\varphi}, \label{dN_over_dpT} 
\end{align}
with $i,j$ indices run over different bottomonium states $\{\Upsilon(1S), \Upsilon(2S), \chi_b(1P), \Upsilon(3S), \chi_b(2P)\}$. After the $b\bar b$ dipoles leave the QGP and the Schr\"odinger evolution ceases, excited bottomonium eigenstates can decay into lower eigenstates, with the branching ratios described by the feed-down matrix ${\bf F}$. The element $F_{ij}$ corresponds to the branching ratio $Br_{j \to i}$ from a higher excited state $j$ to a lower energy bound state $i$. In this work, the information of branching ratios is collected from the Particle Data Group~\cite{ParticleDataGroup:2020ssz}:
\begin{equation}
\boldsymbol{F}=\begin{bmatrix}1&0.2645&0.2475&0.0657&0.0881\\0&0.7355&0&0.1060&0.1245\\0&0&0.7525&0&0.0065\\0&0&0&0.8283&0\\0&0&0&0&0.7809\end{bmatrix},
\end{equation}
where the sum of each column must be equal to unity to preserve the conservation of the bottom number.

In the above equation, $d^2N^{\mathrm{prim}}_{pp,j}/dp_T d\varphi$ is the differential primordial production rate of the $j$th particle in $pp$ collisions, calculated as $d^2N^{\mathrm{prim}}_{pp,j}/dp_T d\varphi = N_{j}^{\mathrm{prim}} f_{pp}(p_T)/(2\pi)$, with $N_{j}^{\mathrm{prim}}$ the integrated primordial production rate, and the normalized initial transverse momentum distribution of the bottomonium bound states can be parameterized as
\begin{align}
f_{pp}(p_T) = \frac{2(n-1)p_T}{(n-2)\langle p_T^2\rangle_{pp}} \left[1 + \frac{p_T^2}{(n-2)\langle p_T^2\rangle_{pp}} \right]^{-n}.\label{eq:f_pp}
\end{align}
In $\sqrt{s}=5.02$ TeV proton-proton collisions,  the parameter is fitted to be $n=2.5$, and the mean square of the transverse momentum of bottomonium ground state $\Upsilon(1S)$ for proton-proton collisions is $\langle p_T^2\rangle_{pp}= 80.0~\mathrm{GeV}^2$. Given that the masses of the bottomonium excited states are close to that of the ground state, this normalized momentum distribution is also applied to the excited states. 

$d^2N^{\mathrm{prim}}_{AA,j}/dp_T d\varphi$ is the number density of each quantum eigenstate produced in a given $p_T$ bin in the final stage of the QGP evolution. It is computed as a convolution of the initial distribution and the transition rate,
\begin{align}
\begin{split}
\frac{d^2N_{AA,n\ell}^{\mathrm{prim}}}{dp_T d\varphi} =\;& 
    \sum_{n^{\prime}}\int d^2 \mathbf{x}\;\rho(\mathbf{x}) \\
    &
    \times \frac{d^2N_{AA,n^{\prime}\ell}^{\mathrm{init}}(\mathbf{x})}{dp_T\,d\varphi}\;P_{n^{\prime}\ell \to n\ell}^{\mathrm{QGP}}(\mathbf{x},p_T, \varphi).
\end{split}
\end{align}
The transition rate $P_{n^{\prime}\ell \to n\ell}^{\mathrm{QGP}}$ is the probability given in Eq.~\eqref{eq:probability_nl} where the initial wave function of $b\bar b$ dipole in terms of relative distance $r$ is initialized with each bottomonium eigenstate $R(t=0,r) = \Psi_{n'\ell}(r)$. The transition rate $P$ depends on the position $x$ and momentum $p_T$ of the initially produced $b\bar{b}$ state.
The distribution of the initial position ($\mathbf{x}$) is given by the binary collision density [$\rho_{\rm bin}(\mathbf{x})$], 
\begin{align}
    \rho(\mathbf{x}) = \frac{\rho_{\mathrm{bin}}(\mathbf{x})}{\int \rho_{\mathrm{bin}}(\mathbf{x}) d^2\mathbf{x}}\,.
\end{align}
They are computed with the Glauber initial condition model, which also provides the initial condition for the hydrodynamic evolution.

The initial (unsuppressed) momentum distribution in AA collisions is given by $d^2N_{n^{\prime}\ell}^{AA,\mathrm{init}}(\mathbf{x}) / dp_T  d\varphi$. It is determined by taking into account the cold nuclear matter effects and the shadowing effect \cite{Chang:2015hqa} on the $pp$ distribution~\eqref{eq:f_pp}, as well as the primordial production rate of bottomonium states in $pp$ collisions~\cite{Islam:2020bnp}. In this work, the shadowing factor of the bottomonium $\mathcal{R}_A(\mathbf{x}_T)$ is calculated using the EPS09 NLO model \cite{Eskola:2009uj}. See, e.g., Ref.~\cite{Wen:2022utn} for more details.
Apparently, $R_{AA,n\ell}(p_T,\varphi)$ depends on the centrality of the collision, since the initial momentum distribution and, more importantly, the hydrodynamic background are centrally dependent.

\subsection{Numerical setting}
Our central computation task is to solve the time-dependent Schr\"odinger equation~\eqref{def_Sch_Eq} numerically. We adopt the Crank--Nicolson formalism~\cite{CrankN96, Chen:2016vha} that discretizes the space-time-dependent wave function on grids and evolve it according to the following formula:
\begin{align}
i\frac{u_j^{n+1}-u_j^n}{\Delta t}=&\frac{1}{2}\left(-\frac{1}{m_b}\frac{u_{j+1}^n-2u_j^n+u_{j-1}^n}{\mathrm{\Delta}r^2}+U_j^n u_j^n\right. \label{eq:discrete_Sch}\\ \notag
&\left.-\frac{1}{m_b}\frac{u_{j+1}^{n+1}-2u_j^{n+1}+u_{j-1}^{n+1}}{\mathrm{\Delta}r^2}+U_j^{n+1} u_j^{n+1}\right),   
\end{align}
where $u_j^{n} \equiv u(n\Delta t,j\Delta r)$ represents the discrete form of the radial wave function $u(t,r)$. $U_j^{n} $ corresponds to the effective potential $U(t,r)$ that includes an angular-momentum contribution in the radial component, i.e.,
\begin{align}
    U(t,r) = \frac{L(L+1)}{m_b r^2} + V(t,r), 
\end{align}
where $V(t,r)$ is the complex potential we discussed in the previous subsection. Traditional techniques usually include the step of finding the inverse matrix, which is time consuming and leads to considerable numerical uncertainty. Thus, a trick named the tridiagonal matrix algorithm (TMA)~\cite{Barata:2025htx}, based on the Gaussian elimination method, is applicable to solve this discrete form of the partial differential equation. The relevant details are available in the Appendix~\ref{sec:app:tma}. Given that the Crank--Nicolson method is unconditionally stable regardless of the discretization sizes of time and position.
Here the discretization conditions are chosen as $\Delta t = 0.1~\textrm{fm}$ and $\Delta r= 0.03~\textrm{fm}$ with a total of $N = 300$ corresponding discrete radial points.
We discretize the momentum direction ($\phi$) in grids $\{\pi/6, \pi/3, \ldots, 2\pi\}$, with the magnitude of the transverse momentum being sampled according to Eq.~\eqref{eq:f_pp}. The initial positions of $b \bar{b}$ dipoles ($\mathbf{x}$) in grids are spaced as $\Delta x =\Delta y = 2~\mathrm{fm}$ within the range $x,y \in [-10,10]~\mathrm{fm}$. 

\subsection{Parameterization of the effective potential}

The real part of the effective bottomonium potential is reduced by color screening in hot plasma. In our study, the slow dissociation limit with a proper modification is selected to describe the bottomonium evolution in the QCD medium. Taking the form of Helmholtz free energy $F$ between a pair of bottom quarks, which can be extracted from lattice QCD calculations~\cite{Kaczmarek:2008saj, Digal:2005ht}, we express the real part of the effective potential in the following formula:
\begin{align}
V_R & = F(r,t) = -\frac{\alpha_s} {r}e^{-m_Dr}+\frac\sigma{m_D}(1-e^{-m_Dr}), \label{ReV}   
\end{align}
where $m_{D}= a_m T \sqrt{\frac{4 \pi N_{c}}{3} \alpha_s(1+\frac{N_{f}}{6})}$ corresponds to the in-medium gluon Debye mass. By solving the reduced Dirac equation with the vacuum Cornell potential and making a comparison with the masses of the bottomonium bound states ($1S$, $1P$), the Coulomb coupling constant and the linear coupling constant can be determined, $\alpha_s=\pi / 12$ and $\sigma=0.2~\mathrm{GeV}^{2}$ with the corresponding vacuum masses of the states being $m_{1S, 1P}=(9.47,9.79)~\mathrm{GeV}$. Here the color and flavor factors are taken as $N_{c}=N_{f}=3$, and the factor $a_m$, considered the first parameter, where the range is chosen as $a_m \in [0,3]$, is used to demonstrate the contribution of the color screening effect on the Deybe mass. Given that both the fit of the Debye mass prefactor to the hard thermal loop (HTL), which yields $a_m\simeq 0.81$ to account for the QGP with all massless $u$, $d$, and $s$ quark degrees of freedom \cite{Kurkela:2018oqw, Du:2020dvp}. Additionally, the fixed coupling constant $\alpha_s$ describes a weak but non-negligible coupling strength, and thus the effects of the strongly coupled medium are to a certain extent absorbed into the $a_m$ parameter in the Bayesian analysis.

To depict the dissociation including the transition from color-singlet to color-octet, which corresponds to the decrease of the inner product of the singlet wave function $\langle \Psi| \Psi \rangle$, the imaginary potential is introduced to describe the contribution of parton inelastic scatterings. We represent the imaginary part of the effective potential using the following formula which incorporates the $r$ and $T$ dependence:
\begin{align}
 V_I(r,T) / m_b = \bar{T}^{f_T} (f_1 \bar{r} + f_2 \bar{r}^{f_3}), \label{ImV}
\end{align}
where $f_T,f_1,f_2,f_3$ are all parameters to be constrained, $f_T \in \left[0.5,3\right]$, $f_1,f_2 \in \left[0,1.0\right]$, $ f_3 \in  \left[1.5,3\right]$. $\bar{r}\equiv r/\mathrm{fm}$ is the scaled radius describing the spatial dependence of the interaction, and $\bar{T}\equiv T/m_b$ is the scaled temperature, where the mass of the bottom quark $m_b$ ensures that the argument of the power is dimensionless. In fact, the artificial specification of parameter ranges here serves as a truncation to facilitate numerical sampling. Nevertheless, the ranges of the five aforementioned parameters are set such that the reconstructed interaction potential can fully cover the observables and their corresponding uncertainties, while ensuring that the range of target parameters derived from the posterior probability falls within this predefined interval, as the Fig.~3 did in the supplementary information in Ref.~\cite{Bernhard:2019bmu}. In the formula for the imaginary part, the contribution of the linear term is controlled by the prefactor $f_1$, with parameters $f_2$ and $f_3$ curving the shape of the effective potential simultaneously.

\section{Bayesian Analysis of the Potential Parameters} \label{sec:bayesian}

In Sec.~\ref{sec:schroedinger}, a quantum time-dependent evolution for the bottominium has already been well established, enabling us to further understand the in-medium effective interaction by determining the parameterized potential. Consequently, a proper data-driven framework should be introduced to explore the credible intervals of the parameter to reconstruct the potential discussed above. In this section, the details of Bayesian estimation (one of the techniques for solving the inverse problems) will be given, followed by the basic procedures for sampling the parameter and a closure test for the validity of the Bayesian framework.

\subsection{Framework of Bayesian estimation}
Generally, the estimation of the parameters based on Bayesian analysis (BA) requires a vector $\boldsymbol{x}$ that represents all parameters to be estimated, as well as its corresponding model output $\boldsymbol{y}(\boldsymbol{x})$, which is another vector representing the physical observables to be compared with experimental data ($\boldsymbol{y}_{\rm exp}$). By evaluating the difference between $\boldsymbol{y}_{\rm exp}$ and $\boldsymbol{y}(\boldsymbol{x})$, BA provides the likelihood distribution of the parameters derived from Bayes's theorem,
\begin{align}
    \mathcal{P}_{\mathrm{posterior}}(\boldsymbol{x}|\boldsymbol{y}_{\mathrm{exp}}) = \frac{\mathcal{P}(\boldsymbol{y}_{\mathrm{exp}}|\boldsymbol{y}(\boldsymbol{x})) \mathcal{P}_{\mathrm{prior}}(\boldsymbol{x})}{\mathcal{P}(\boldsymbol{y}_{\mathrm{exp}})},
\end{align}
further determines the credible intervals, and the median value for the Bayesian parameters can then be determined. $\mathcal{P}(\boldsymbol{y}_{\mathrm{exp}})$ is the constant that ensures the normalization of the posterior distribution $1=\int \mathcal{P}_{\mathrm{posterior}}(\boldsymbol{x}) d^d\boldsymbol{x}$. It is also called the ``Bayesian model evidence" and describes the validity of our model given the experimental observables.
In practice, we take the parameters as $\boldsymbol{x} \equiv \{a_m, f_T, f_1, f_2, f_3\}$, and $\boldsymbol{y}$ is the set consisting of $R_{AA}$'s for various bottomonium states at different centrality bins and for different transverse momentum.

In the above equation, $\mathcal{P}_{\mathrm{prior}}(\boldsymbol{x})$ is the prior distribution that encodes our estimation of the parameter distribution without comparison with the data. As a standard approach (see, e.g., Ref.~\cite{Bernhard:2019bmu}), we assume no extra knowledge of the parameters other than the range they belong to, as discussed above.
Thus, we adopt a multidimensional uniform distribution within the corresponding intervals:
\begin{align}
\mathcal{P}_{\mathrm{prior}}(\boldsymbol{x})= 
    \prod_{i=1}^{d}\frac{\Theta((x_{\mathrm{max}}^{(i)}-x^{(i)})(x_{}^{(i)}-x_{\mathrm{min}}^{(i)}))}{(x_{\mathrm{max}}^{(i)}-x_{\mathrm{min}}^{(i)})} 
.
\end{align}
That is, the value of the prior probability is a constant, enabling us to apply the Latin-hypercube method to the subsequent process of producing mock data.

$\mathcal{P}(\boldsymbol{y}_{\mathrm{exp}}|\boldsymbol{y}(\boldsymbol{x}))$ is the probability of the experimental data given the prediction of the model under the set of parameters $\boldsymbol{x}$. It generally follows the multivariate normal distribution with variance given by both theoretical and experimental uncertainties:
\begin{align}
\mathcal{P}(\boldsymbol{y}_{\mathrm{exp}}|\boldsymbol{y}(\boldsymbol{x}))   
= \frac{\exp\left[-\frac{1}{2} (\Delta \boldsymbol{y}_{\boldsymbol{x}})^{T} \cdot\Sigma^{-1}(\boldsymbol{x}) \cdot \Delta \boldsymbol{y}_{\boldsymbol{x}}\right]}{\sqrt{(2\pi)^{d} \det\left[\Sigma(\boldsymbol{x}) \right]}}.
\label{eq:likelihood}
\end{align}
Here $\Delta \boldsymbol{y}_{\boldsymbol{x}} \equiv \boldsymbol{y}(\boldsymbol{x}) - \boldsymbol{y}_{\mathrm{exp}}$ is defined as the deviation between theoretical and experimental values, which is represented as a vector of dimension $d$.
The prediction of the model $\boldsymbol{y}(\boldsymbol{x})$ is calculated, in principle, by numerically solving the Schr\"odinger equation as discussed in the preceding subsection. However, this process is time-consuming, and for each switching temperature ($T_d$) we perform a numerical solution for 200 designed points in the five-dimensional Latin hypercube~\cite{tang1993orthogonal}. We then perform principal component analysis (PCA) because the responses of the observables' to the parameters can be cast into a few principal components (PCs). We choose five PCs in this work, and we train a Gaussian processes (GP) emulator separately for each independent PC dimension to obtain $\boldsymbol{y}(\boldsymbol{x})$ in the full space. The scikit-learn library in Python is utilized to construct GP for each PC, where the self-optimization process of parameters employs Hamiltonian Monte Carlo sampling and the maximum logarithmic marginal likelihood estimation. More details can be found in Appendixes~\ref{chapterGP} and~\ref{SectionPCA}, respectively, for GP and PCA.

To quantify the uncertainty of both the experimental measurements and theoretical predictions, we construct the covariance matrix $\Sigma(\boldsymbol{x})$ that is regulated by the parameter $\boldsymbol{x}$, whose definition can be expressed by the following formula:
\begin{align}
    \Sigma(\boldsymbol{x}) &= \Sigma_{\mathrm{th}}(\boldsymbol{x}) + \Sigma_{\mathrm{exp}}, \\
    \Sigma_{\mathrm{exp}} &= \Sigma_{\mathrm{exp}}^{\mathrm{stat}} + \Sigma_{\mathrm{exp}}^{\mathrm{sys}},\\
    \Sigma_{\mathrm{exp}}^{\mathrm{stat}} &= \mathrm{diag} \left(\delta_{\mathrm{stat},1}^2, \delta_{\mathrm{stat},2}^2, \cdots, \delta_{\mathrm{stat},d}^2 \right),   
\end{align}
where the experimental covariance matrix encodes both statistical and systematic uncertainties, with the former being representable as a diagonal matrix. Systematic uncertainties are generally correlated with each other. We have performed the analysis with the full covariance matrix based on the correlations of the available experimental data and found the  results indistinguishable with those only considering the diagonal systematic uncertainties (see Appendix~\ref{Sigma_syst}). Therefore, a diagonal systematic covariance matrix is utilized in the main text. See, e.g., Refs.~\cite{JETSCAPE:2021ehl, JETSCAPE:2024cqe} for more discussions of estimating the effect of nondiagonal terms in other contexts.
The theoretical covariance matrix $\Sigma_{\mathrm{th}}$, on the other hand, is not diagonal. It is constructed by applying the PCA's decoding matrix to the diagonal covariance matrix in the PC's space.

\subsection{Parameter sampling} \label{ChapterMCMC}

With the GP serving as a fast surrogate for the solver of the time-dependent Schr\"odinger equation, we may move on to a massive sampling of the parameters according to the posterior. This is usually done based on the Markov-chain Monte Carlo (MCMC) method, which derives from the stationary distribution of the rejection sampling method. From this distribution, the target set of parameters can then be collected to reconstruct the posterior. The relevant details of the MCMC are available in Appendix~\ref{Hastings}. 

We invoke the Hamiltonian Monte Carlo (HMC)~\cite{betancourt2018}, a gradient-based improvement of the basic MCMC method, to sample parameter sets according to the posterior distribution $\mathcal{P}_{\mathrm{posterior}}(\boldsymbol{x}|\boldsymbol{y}_{\mathrm{exp}})$, in a computationally efficient manner. HMC accumulatively samples a chain of parameter sets,
$\{ \boldsymbol{x_1}, \boldsymbol{x_2}, \boldsymbol{x_3}, \ldots \}$, by repeating the iterations described in Appendix~\ref{HamiltonianMonteCarlo}.
With sufficient parameter sets collected, their distributions converge to the posterior distribution.

\begin{figure}[htbp!]
    \centering
        \centering
        \includegraphics[width=.45\textwidth]{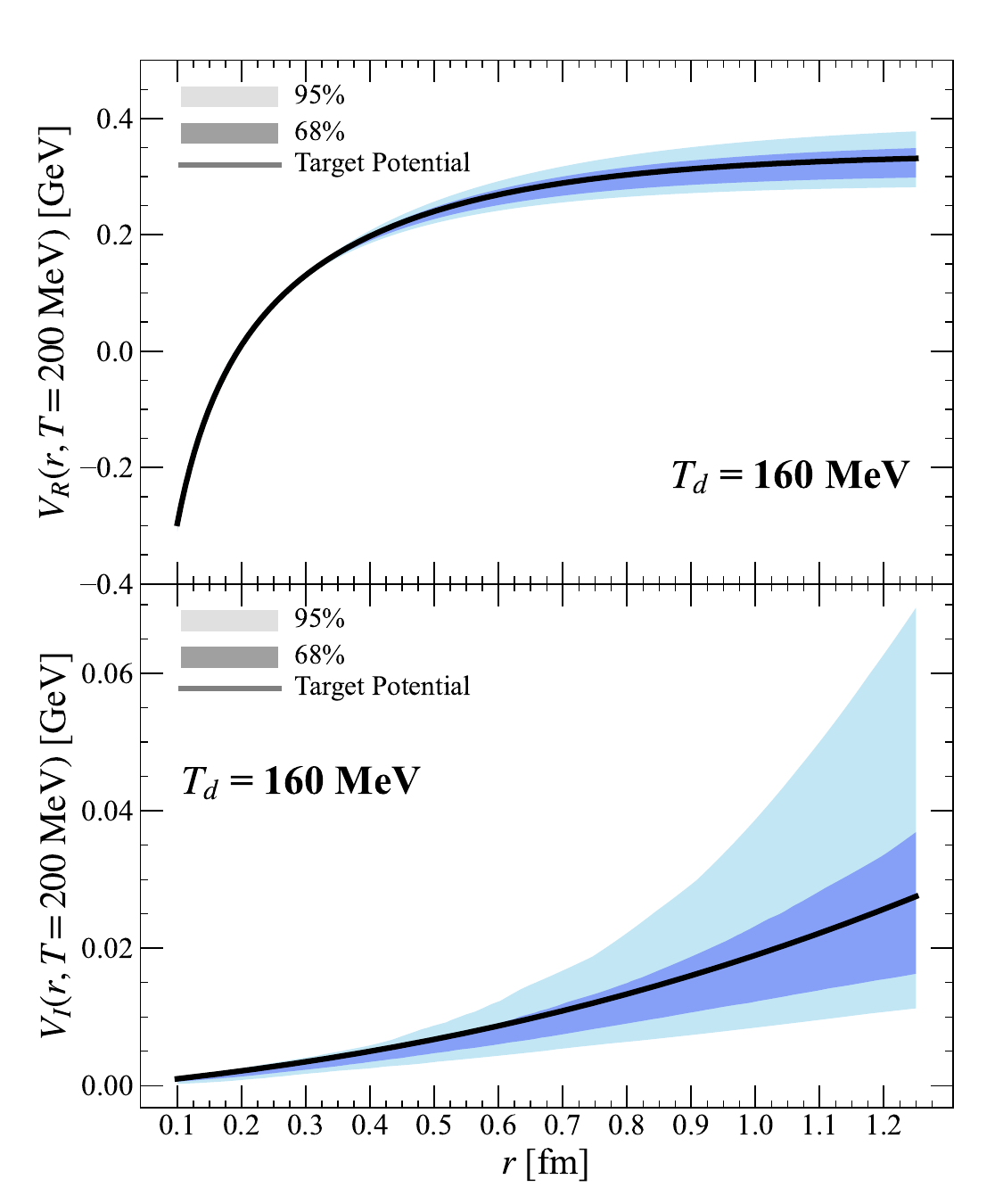}
    \caption{The comparisons between the results of closure test and the given analytical potential where $T_d = 160$ MeV, including the real (upper) and the imaginary part (lower) as functions of the radius $r$, where the temperature is chosen as $T=200$ MeV. The $68\%$ and $95\%$ creditable intervals are represented as the light blue and blue bands, respectively. The black-solid lines represent the mock, analytical potential corresponding to the target median parameter. The parameter for mock test come from the MCMC sampling, where the selection of the emulators is the same as that of the real estimation.}
    \label{fig:MockTest}
\end{figure}
Here $-2\ln \mathcal{P}_{\mathrm{posterior}}(\boldsymbol{x}|\boldsymbol{y}_{\mathrm{exp}})$ can be treated as a loss function when optimizing the parameters, and it returns to the traditional $\chi^2$ ~\cite{Du:2019tjf} when the prior distribution is flat. And the $\chi^2$ fit returns in the ``most optimal'' parameter set that minimizes the loss function or, equivalently, maximizes the posterior distribution. In a BA that aims for a probabilistic description of the theoretical model, the intervals of the target parameters are distributed around such an ``optimal'' parameter set. However, the volume of this point in phase space is so small that it cannot be reproduced efficiently. BAs usually adopt the posterior median in parameter estimation. Credible intervals for parameters are then obtained from the cumulative distribution.

\subsection{Closure test}
To ensure that our BA framework is capable of correctly predicting the actual effective potential of their corresponding experimental observables \cite{JETSCAPE:2020mzn}, we perform a closure test as presented in this subsection. Ideally, the posterior of a closure test with vanishing uncertainties should take the form of a $\delta$ function, that is,
\begin{equation}
\lim_{\delta\to0}\mathcal{P}_{\mathrm{posterior}}(\boldsymbol{x}| \hat{\boldsymbol{y}}_{\mathrm{mock}}) = \delta(\boldsymbol{x}- \hat{\boldsymbol{x}}_{\mathrm{mock}}),
\end{equation}
where the mock observables $\hat{\boldsymbol{y}}_{\mathrm{mock}}$ correspond to the mock selection of a parameter specifically chosen $\hat{\boldsymbol{x}}_{\mathrm{mock}}$ that can reconstruct the given effective potential $V(r,T | \hat{\boldsymbol{x}}_{\mathrm{mock}})$. Nevertheless, the uncertainty caused by both experimental statistics and theoretical emulators will smear the ideal distribution. Therefore, the expectation of the mock test should be displayed in the form of a distribution for the effective potential, containing information about a credible interval (CI).

In this mock test, without loss of generality, we choose a parameter set as $\hat{\boldsymbol{x}}_{\mathrm{mock}} =(\hat{a}_m, \hat{f}_T, \hat{f}_1, \hat{f}_2, \hat{f}_3)^T = (1.5, 1.75,0.5,0.5,2.25)^T$ and select the switching temperature for the bottomonium evolution to be $T_d = 160~\mathrm{MeV}$.

By numerically solving the Schr\"odinger equation with the ``mock potential'' and treating the model output of $R_{AA}$'s as ``mock data'' (with uncertainties taken from actual experiments), we perform the BA by accumulating a total of 800 parameter points from eight different Markov chains. To quantify the possible difference between the mock potential and that given by the BA, we compare the potentials at a fixed temperature $T = 200~\mathrm{MeV}$, such that the potential becomes a single-variable function of the distance ($r$).

The results of the closure test are shown in Fig.~\ref{fig:MockTest}, where the mock potential is represented by the black solid lines. Meanwhile, the $68\%$ and $95\%$ CIs of BA are shown as darker and lighter blue bands, respectively. They recover the mock potential, which validates the Bayesian framework. We also note that the reconstructed potential exhibits credible intervals due to experimental uncertainties, especially in the large distance range $r > 1.2~\mathrm{fm}$. The latter is well expected because the bottomonium states of interest typically have a radius smaller than, or similar to, $1~\mathrm{fm}$. Thus, their production is not sensitive to the potential at a distance larger than that.

\begin{figure}[htbp!]
    \centering
    \includegraphics[width=.45\textwidth]{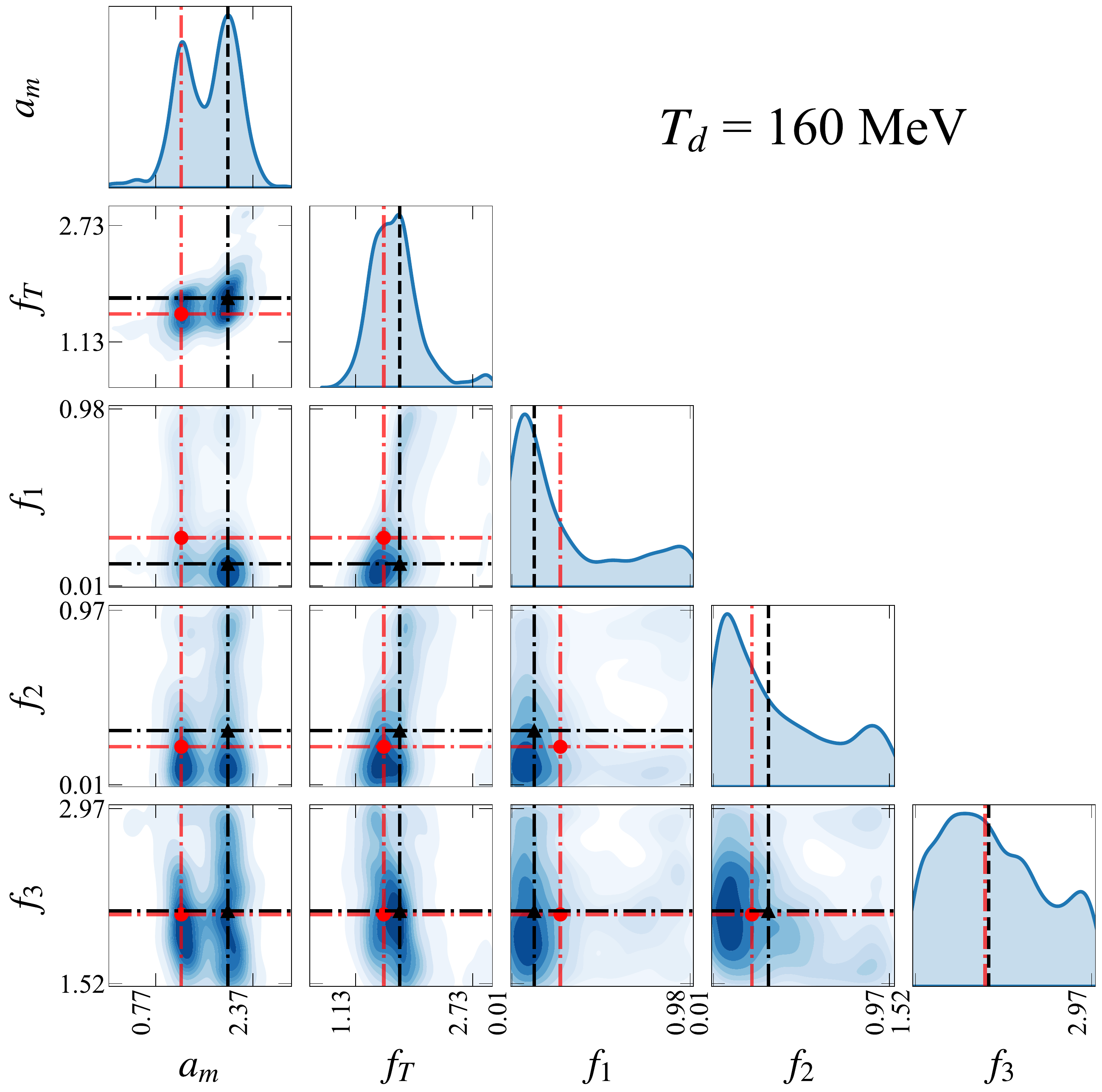}
    \caption{The posterior distributions of five parameters of Bayesian analysis for the switching temperature $T_d = 160~\mathrm{MeV}$. The histograms on the diagonal are the marginal distributions for each parameter, and the off-diagonal subgraphs correspond to the 2D joint probability distributions of parameter pairs.
    Noting the clear bimodal distribution, we show median parameters separately for two peaks, indicated by red and black dash-dot lines, respectively.}
    \label{fig:Posterior_160}
\end{figure}

\section{Results} \label{sec:results}
With the frameworks for solving $b\bar{b}$-state evolution and Bayesian inference established and tested, we are now ready to perform BA of the bottomonium potential against the experimental data of $1S$, $2S$, and $3S$ $\Upsilon$ states' $R_{AA}$ at various centrality and transverse momentum ($p_T$) bins, measured by ALICE \cite{ALICE:2018wzm}, ATLAS \cite{ATLAS:2022exb}, and CMS \cite{CMS:2018zza,CMS:2023lfu} Collaborations, respectively, for $\sqrt{s_{NN}}=5.02~\mathrm{TeV}$ Pb+Pb collisions.
\begin{figure}[!htbp]
    \centering
        \includegraphics[width=.45\textwidth]{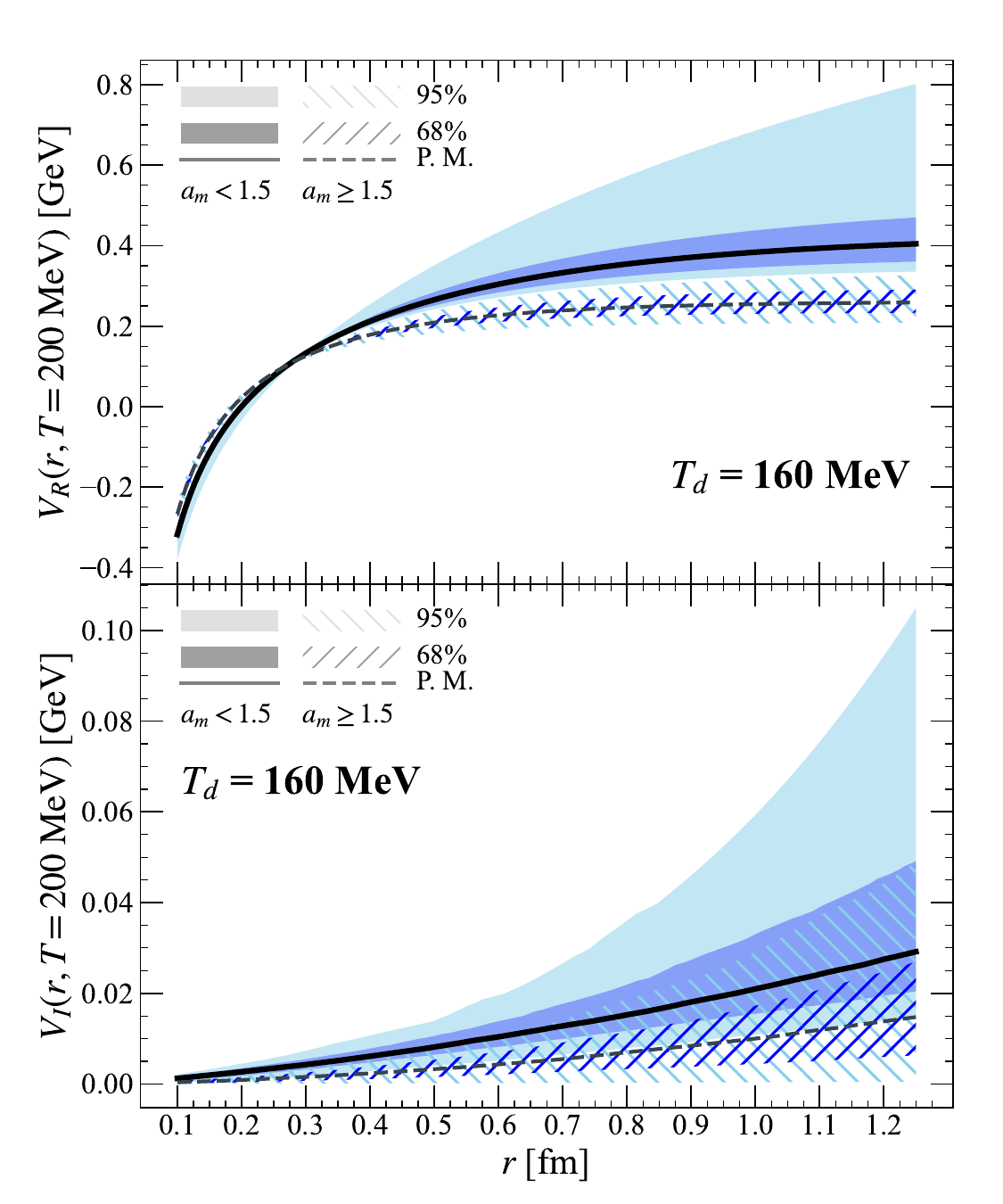}
    \caption{The $68\%$ (darker blue) and $95\%$ (lighter blue) creditable intervals and median values (black lines) of the effective potential as functions of radius $r$ for temperature $T = 200~\mathrm{MeV}$. The two peaks in the bimodal posterior distribution are show separately with filled bands ($a_m < 1.5$) and hatch shading ($a_m \geq 1.5$), respectively. \label{fig:EffectvePotential_160}}
\end{figure}
We begin by setting the switching temperature as $T_d=160~\mathrm{MeV}$ and extracting the potential parameters from BA. The one- and two-dimensional marginal posterior distributions are shown in Fig.~\ref{fig:Posterior_160}. One can observe a bimodal structure, with one peak sizably dominating. A bimodal structure usually indicates that two different parameter sets are comparably favored by the experimental data. The two peaks in the posterior distribution can be well separated for the dressing factor of the Debye screening mass, $a_m$. With peaks centered at approximately $1.19$ and $1.96$, respectively, we observe that the junction between the two modes occurs at $a_m = 1.5$.  Thus, this value is chosen as the threshold for cutting the distribution into two peaks. We found that the $a_m < 1.5$ peak occupies $40.8\%$ of the total likelihood, whereas the $a_m \geq 1.5$ one contributes to the remaining $59.2\%$. The maximal posterior density of the latter is $1.45$ times of that of the former, which makes it more favored according to the BA.

\begin{figure}[!htbp]
    \centering
    \includegraphics[width=0.42\textwidth]{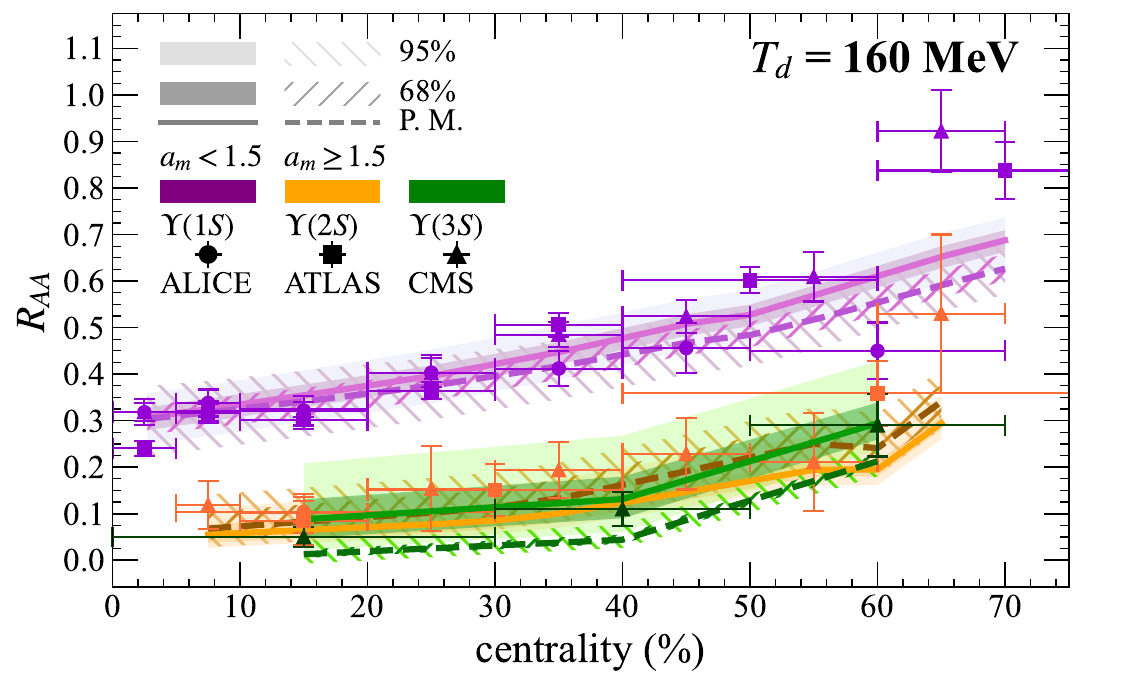}
    \includegraphics[width=0.42\textwidth]{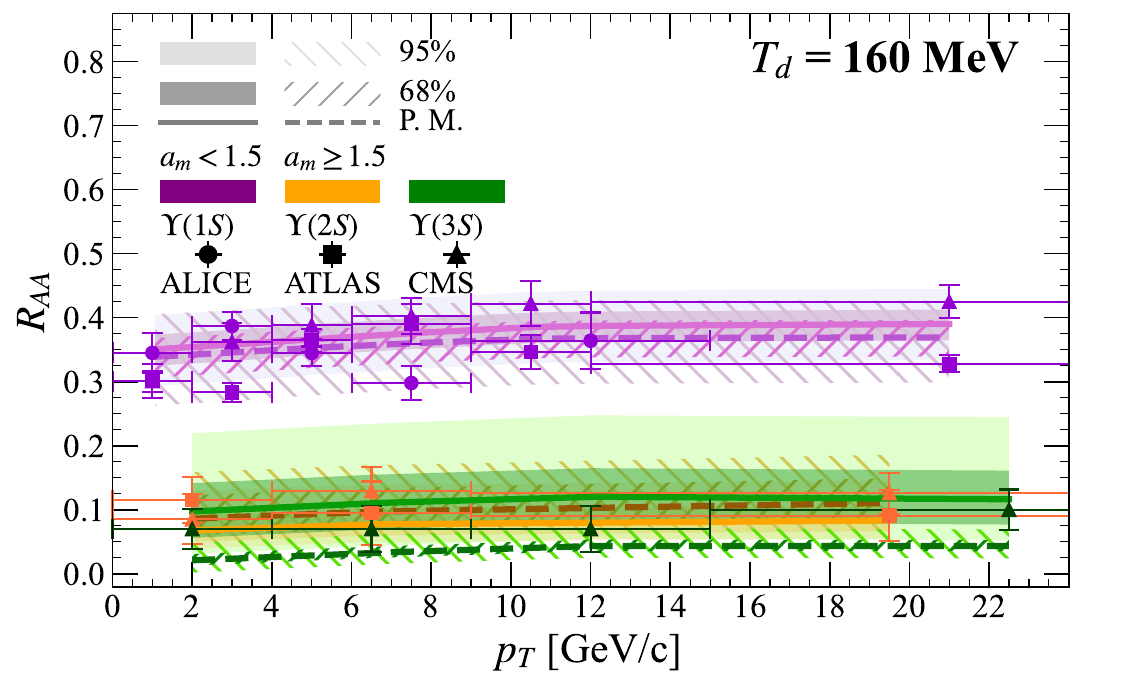}
    \caption{Nuclear modification factors of $1S$ (purple), $2S$ (orange), and $3S$ (green) states of $\Upsilon$ as functions of centrality (top) and the transverse momentum (bottom). Conventions for peak separation and posterior distribution is the same as in Fig.~\protect{\ref{fig:EffectvePotential_160}}.
    Experimental data from ALICE \cite{ALICE:2018wzm}, ATLAS \cite{ATLAS:2022exb}, and CMS \cite{CMS:2018zza,CMS:2023lfu} Collaborations are also shown from comparison.}
    \label{fig:RAA_160MeV}
\end{figure}
With the posterior distribution separated into two peaks, we may move on to obtain the corresponding potentials.
Figure \ref{fig:EffectvePotential_160} displays the effective potential as a function of radius $r$ at a given finite temperature $T = 200$ MeV. Here the fixed temperature is selected just above the switching temperature $T_d=160$ MeV for both quantifying the effects at the end of in-medium evolutions and setting a standard for comparison with different switching temperatures. The filled bands (hatch shading) represent the scenario where $a_m < 1.5$ ($a_m \geq 1.5$). We note that one of the peaks, $a_m\simeq1.96$, lies around the aforementioned HTL results~\cite{Kurkela:2018oqw, Du:2020dvp}, and the other corresponds to the weakly screening scenario seen in the lattice QCD result~\cite{Bala:2021fkm}. Indeed, such degeneracy coincides with two scenarios of lattice QCD results---the less favored potential, with $a_m < 1.5$, exhibits a mild screening effect in the real potential but a large imaginary potential, while the more favored one ($a_m \geq 1.5$) predicts that the bottomonium states shall be caused mostly by the Debye screening effect rather than by the dissociation including the transition from color-singlet to color-octet states.

Observing such a degeneracy, a natural suspect for its origin is the parameter-dependent theoretical uncertainty caused by the Gaussian emulator. Imagine that one of these peaks corresponds to a region where the theoretical uncertainties are very large, resulting in wide prediction bands that always cover the data. This possibility has been ruled out by a test that fixes $\Sigma_{\mathrm{th}}$ to be constant (by taking the parameter average), in which we observe a quantitatively similar posterior distribution as shown in Fig.~\ref{fig:Posterior_160}.

This nonphysical degeneracy is usually caused by the uncertainties arising from both experimental observations and theoretical calculations. To address the possible local minima when sampling with MCMC, the theoretical uncertainty $\Sigma_{\mathrm{th}}(\boldsymbol{x})$ for each point in the phase space $\boldsymbol{x}$ is equalized, where the covariance for different PC's is selected as the average covariance. However, there is almost no distinction between cases using different theoretical uncertainties and those using the same uncertainty.

To better trace the origin of the weak degeneracy and find routes to resolve it, we compute the modification factors ($R_{AA}$) separately for $V(r,T)$ described by the two peaks in the posterior distributions, with results presented in Fig.~\ref{fig:RAA_160MeV}.
In the centrality dependent plot, we focus on the transverse momentum region $p_T < 30~\mathrm{GeV}/c$, whereas the $p_T$ dependence is for the $0$--$80\%$ centrality class.
Given the width in $P_{\mathrm{posterior}}$, the corresponding $R_{AA}$ exhibits sizable uncertainties, and one cannot distinguish the two sets of parameters beyond uncertainty bands---both in broad agreement with the experimental results.
Upon a close look at the $R_{AA}$ of $2S$ and $3S$ states, we observe that the strong-screening case ($a_m \geq 1.5$) follows the trend of sequential suppression that $R_{AA}(2S) > R_{AA}(3S)$~\cite{Digal:2001ue}.
For the weak-screening ($a_m < 1.5$) case, case, the effect of sequential suppression is not as pronounced as in the former case, because decay rates for $\Upsilon(2S)$ and $\Upsilon(3S)$ are comparable in magnitude. Thus, we anticipate that a precise measurement of the excited states could eventually resolve the degeneracy. Additionally, the reconstructed results for the elliptic flow $v_2$ indicate that current experimental measurements cannot lift the degeneracy of the potential, as their magnitudes are too small relative to the experimental uncertainties.

\vspace{5mm}
While the pseudocritical temperature, $T_{\mathrm{pc}} = 160~\mathrm{MeV}$, seems to be a natural choice for the switching temperature, the crossover nature of the phase transition allows for a different choice of $T_d$. Noting the hadron gas component above $T_{\mathrm{pc}}$ which suppresses the bottomonium-QGP interaction, in what follows, we test the possibilities of setting $T_{d}$ to be slightly above $T_{\mathrm{pc}}$. We respectively take $T_d = 180~\mathrm{MeV}$ and $T_d = 200~\mathrm{MeV}$ and repeat the aforementioned BA separately.

\begin{figure*}[htbp!]\centering
    \includegraphics[width=0.45\textwidth]{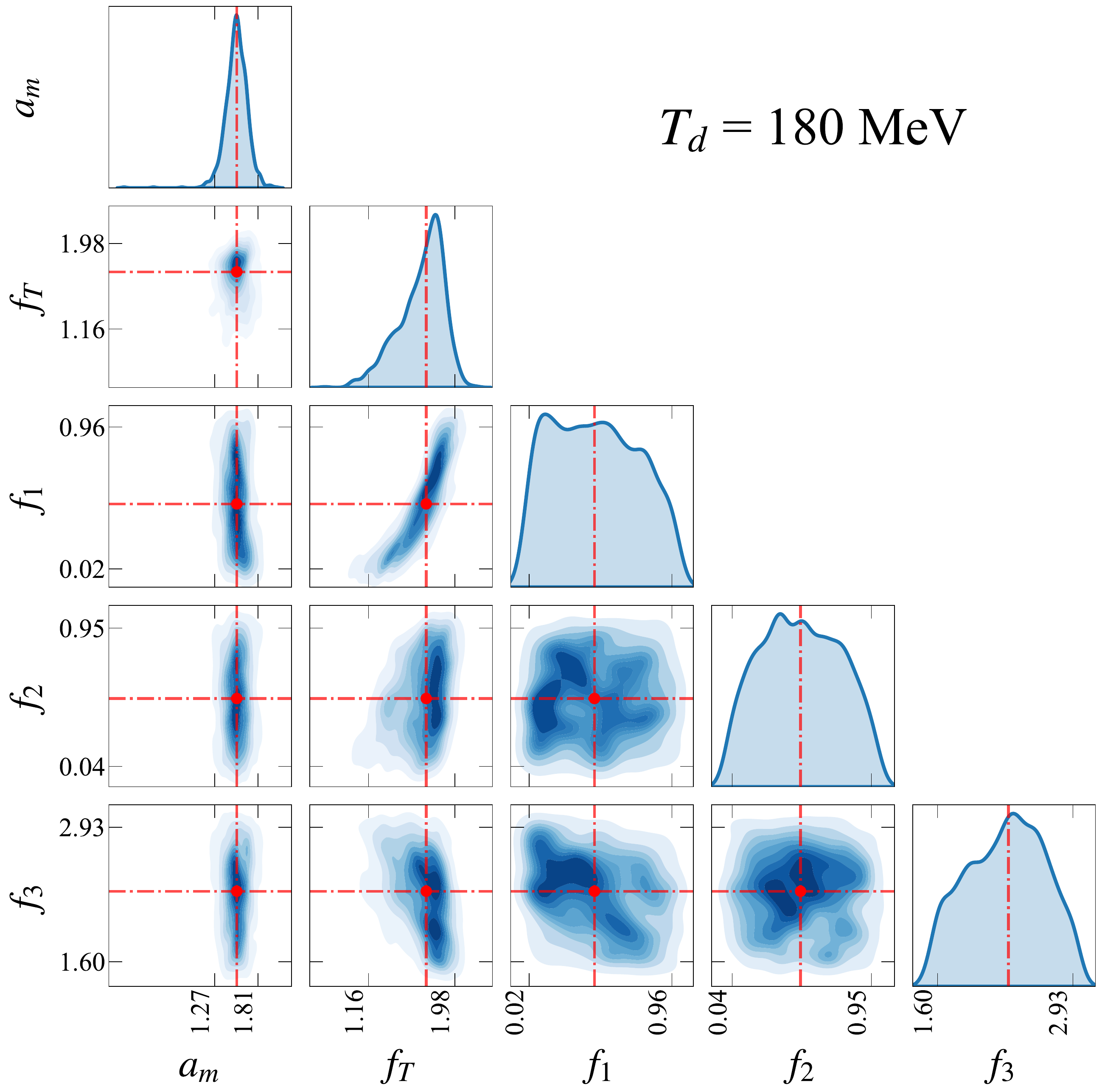}
    \includegraphics[width=0.45\textwidth]{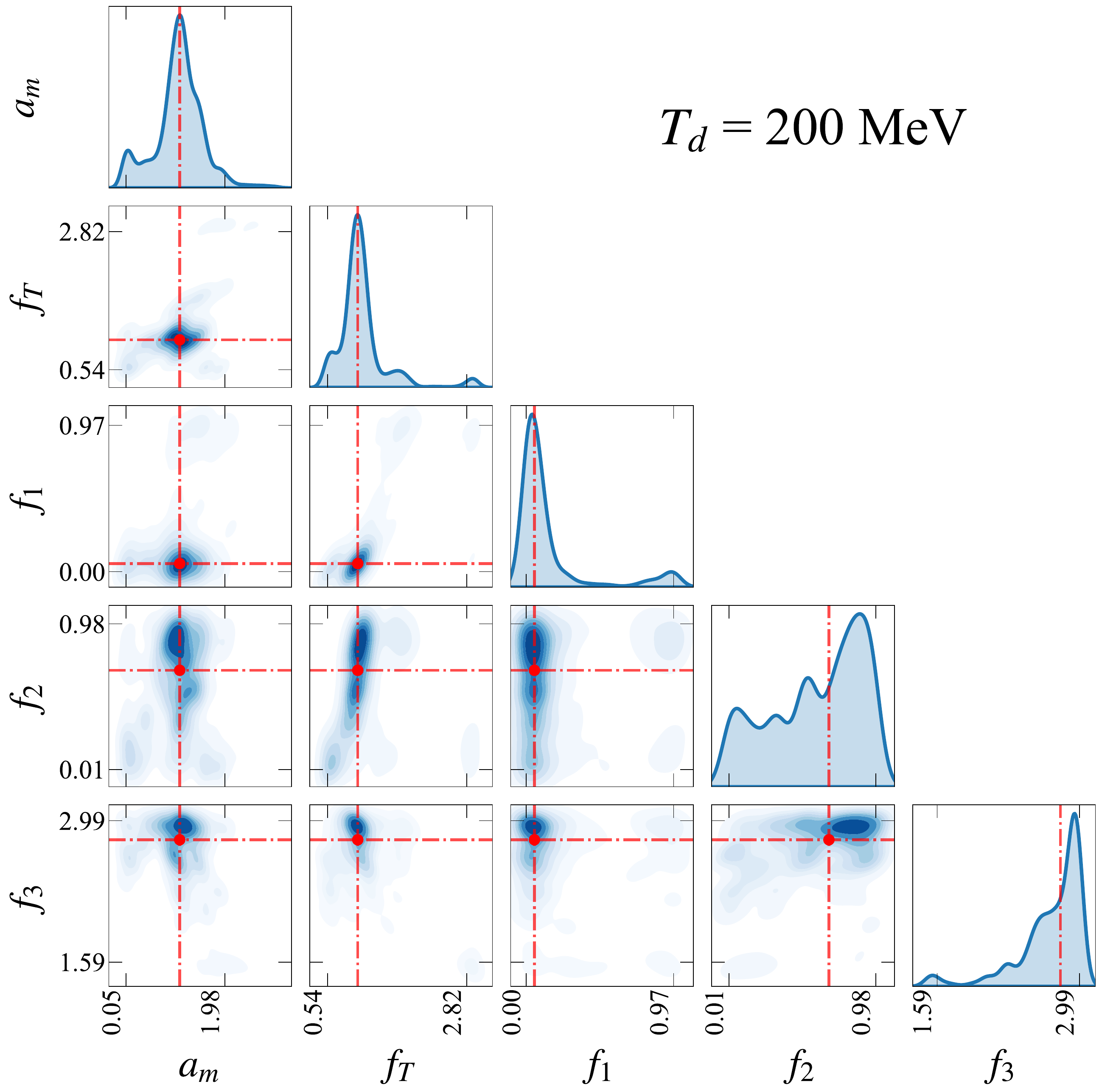}
    \caption{Same as Fig.~\ref{fig:Posterior_160} but for switching temperatures set to be $T_d = 180~\mathrm{MeV}$ (left) and $T_d = 200~\mathrm{MeV}$ (right), respectively.
    \label{fig:Posterior_higher}}
\end{figure*}

\begin{figure}[htbp!]
    \centering
        \includegraphics[width=0.45\textwidth]{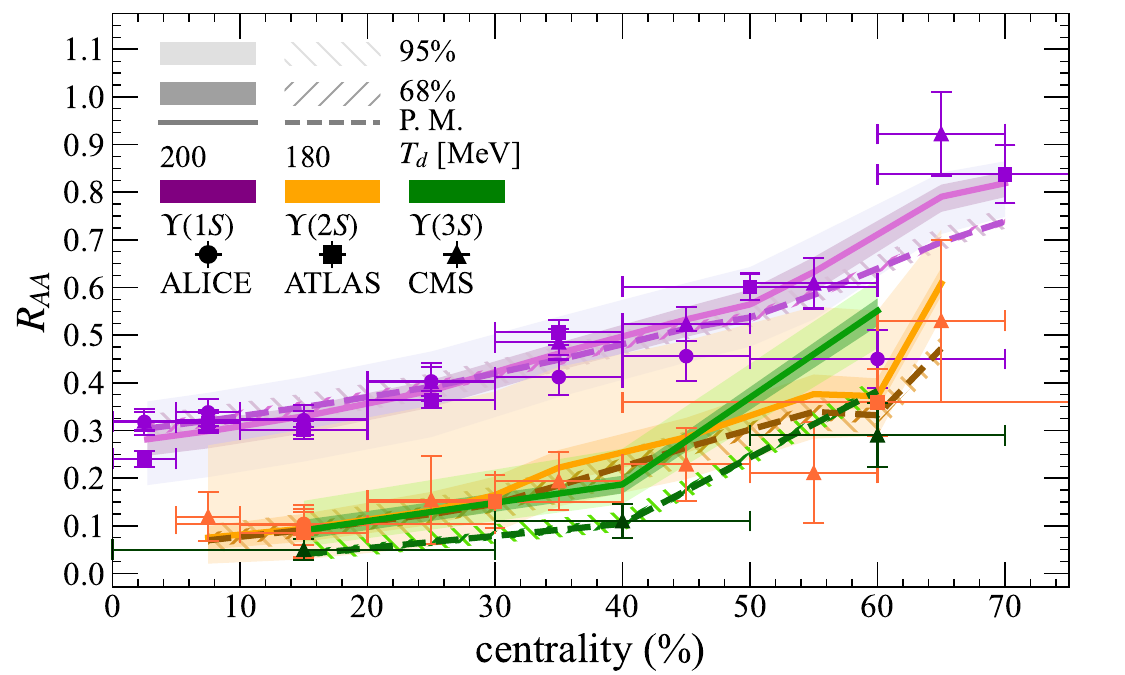}
        \includegraphics[width=0.45\textwidth]{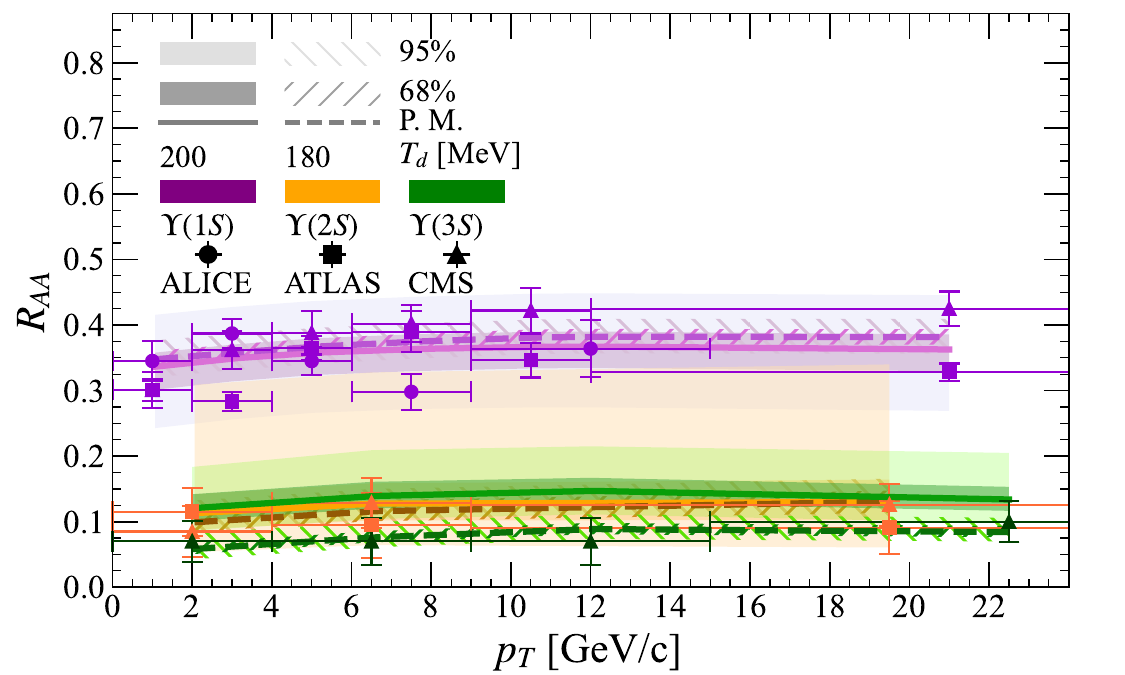}
    \caption{Same as \protect{Fig.~\ref{fig:RAA_160MeV}} but for $T_d = 200~\mathrm{MeV}$ (filled bands) and $T_d = 180~\mathrm{MeV}$ (hatch shading).
    \label{fig:RAA_higher}}
\end{figure}
Resulting posterior distributions of the potential parameters are shown in Fig.~\ref{fig:Posterior_higher}, for the two choices of switching temperature, respectively. Interestingly, the posterior exhibits a clear monomodal structure in both cases, and no degeneracy has been observed. while the degeneracy still exists for the other case. As one can observe in Fig.~\ref{fig:RAA_higher}, The calculated $R_{AA}$ agrees well with the LHC data for $T_d = 180~\mathrm{MeV}$ and for the $p_T$ dependence. Nevertheless, in the centrality dependent $R_{AA}$ for $T_d = 200~\mathrm{MeV}$, the yield of the $2S$ state is noticeably lower than that of the $3S$ state in peripheral collisions, which is not exhibited in the experimental data. Therefore, within the framework of the present model based on the time-dependent Schr\"{o}dinger equation, $T_d = 200~\mathrm{MeV}$ does not appear to be supported as the switching condition from in-medium to vacuum interactions.

\begin{figure}[htbp!]
    \centering
        \includegraphics[width=0.45\textwidth]{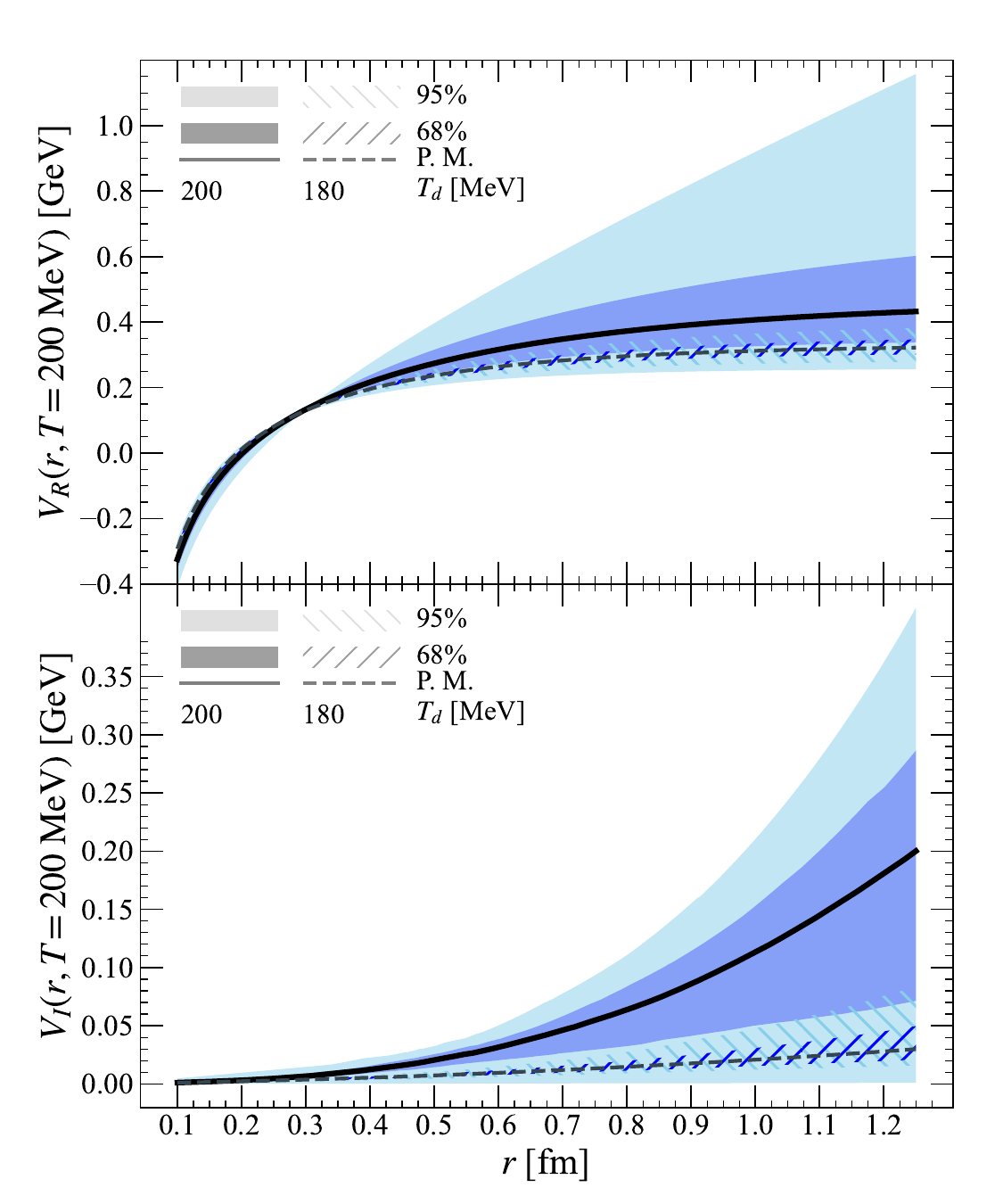}
    \caption{Same as \protect{Fig.~\ref{fig:EffectvePotential_160}} but for $T_d = 200~\mathrm{MeV}$ (filled bands) and $T_d = 180~\mathrm{MeV}$ (hatch shading).
    \label{fig:EffectvePotential_higher}}
\end{figure}

In the distribution of the Debye screening mass coefficient $a_m$, we observe that the peak for the case where $T_d = 200~\mathrm{MeV}$ is close to the left one (dominated by the color-octet transition) of the bimodal structure displayed in Fig.~\ref{fig:Posterior_160}, while that for $T_d = 180~\mathrm{MeV}$ shows similarity to the right peak corresponding to the screening-dominant ($a_m \geq 1.5$) peak.
Such a correspondence can also be observed in the explicit form of the posterior potential, e.g., the $r$ dependence with temperature fixed at $T=200~\mathrm{MeV}$ as presented in Fig.~\ref{fig:EffectvePotential_higher}.

Note that a higher switching temperature always corresponds to a shorter evolution time and results in less suppression given the same potential. Thus, with higher $T_d$, potentials with stronger screening and/or more dissociation including the transition to color-octet are needed to reproduce the same $R_{AA}$'s. One can observe this from a detailed comparison between Figs.~\ref{fig:EffectvePotential_160} and \ref{fig:EffectvePotential_higher}. Let us focus on the values at the largest distance ($r = 1.25~\mathrm{fm}$). From the dashed lines, we read $V_R = 259^{+31}_{-26}~\mathrm{MeV}$ and $V_I = 15^{+12}_{-9}~\mathrm{MeV}$ in Fig.~\ref{fig:EffectvePotential_160}, while such values become $V_R = 321^{+26}_{-20}~\mathrm{MeV}$ and $V_I = 30^{+20}_{-9}~\mathrm{MeV}$ in Fig.~\ref{fig:EffectvePotential_higher}, which correspond to slightly weaker screening and a stronger color-octet transition in the latter.
Likewise, for the solid curves, we read $(V_R, V_I) = (404^{+66}_{-44}, 29^{+20}_{-9})~\mathrm{MeV}$ in Fig.~\ref{fig:EffectvePotential_160} and $(432^{+170}_{-94}, 200^{+86}_{-129})~\mathrm{MeV}$ in Fig.~\ref{fig:EffectvePotential_higher}, which exhibits similar trends with the aforementioned strong screening scenario.
All numbers quoted here correspond to the median and the $68\%$ CI.

\section{Summary and Outlooks} \label{sec:summary}
In this work, we perform a data-driven analysis to constrain the effective interaction potential of bottomonium in the hot QCD medium. 
We adopt a parametrization of the potential that incorporates a tunable color screening effect in the real part, along with an undetermined strength including the color-singlet to color-octet transition characterized by the imaginary part.
By numerically solving the nonrelativistic time-dependent Schr\"odinger equation, we establish the relationship between the potential parameters and key experimental observables such as the nuclear modification factor ($R_{AA}$). We then perform a Bayesian analysis of the potential parameters using experimental measurements of different $\Upsilon$ states' $R_{AA}$.

Our Bayesian framework incorporates Latin-hypercube sampling for parameter space exploration, Gaussian emulators as fast surrogates for simulations, and Hamiltonian Monte Carlo sampling for efficient estimation of the posterior distribution of parameters, ensuring robust extraction of the potential.
A closure test validates the reliability of our framework, demonstrating that it can effectively reconstruct the input mock potential within reasonable uncertainty bounds, accounting for both experimental and theoretical uncertainties. 

Our Bayesian analysis results in two distinct bottomonium potentials. It favors that the suppression is dominated by the screening of the real part, whereas the alternative, which predicts a large imaginary part corresponding to the color-singlet to color-octet transition, cannot be excluded. We anticipate that such a degeneracy will be lifted if the experimental uncertainties in the $R_{AA}$ for the excited states are reduced.

While these results correspond to adopting the switching temperature as $T_d = T_{\mathrm{pc}} = 160~\mathrm{MeV}$, we also investigate the effect of higher values of $T_d$. Our Bayesian analyses with $T_d=180~\mathrm{MeV}$ and $T_d=200~\mathrm{MeV}$, seem to resolve the degeneracy, but they respectively align with the two degenerate scenarios and do not yield a consistent physical picture.

A natural question is why two well separated potentials are found from the Bayesian analysis, rather than a continuous trajectory connecting them. We conjecture that this may stem from the power form of the parametrization of the potentials---other points of the aforementioned trajectory do not fall within the scope of the current parametrization scheme. Nevertheless, identifying such a trajectory help lift the degeneracy and is therefore beyond the scope of this work. We anticipate that more accurate measurements of the bottomonium states' in-medium suppression, especially the high-statistics Au+Au collisions with $\sqrt{s_{NN}} = 200~\mathrm{GeV}$, will ultimately enable a definitive determination of the bottomonium potential.

It shall also be worth noting the quantitative comparison between our extracted potential and the lattice QCD calculations~\cite{Rothkopf:2011db, Burnier:2014ssa, Burnier:2015tda, Bala:2019cqu, Bala:2021fkm, Bazavov:2023dci, Ali:2025iux}. As seen in~\cite{Chen:2024iil}, performing the same calculation with lattice QCD results~\cite{Bala:2021fkm} would predict significantly more suppression than the experiment results. Therefore, our theory framework of bottomonium evolution prefers less screening in the real potential and less transition from color-singlet to color-octet.
Such a tension could probably be resolved by taking into account the recombination of dissociated $b\bar{b}$ pairs --- a recent phenomenological estimation~\cite{Wu:2025lcj} indicates that the ``free'' $b$ and $\bar{b}$ pair dissociated from a bottomonium can reform a color-singlet boundstate during the evolution and eventually reach a equilibrium. We will address this issue in the future.

\section*{Acknowledgments}
We thank Dr. Swagato Mukherjee and Dr. Alexander Rothkopf for useful discussions, and Dr. Yu Guo and Haiyang Shao for their helpful technical support. This work is supported by Tsinghua University under Grants No. 04200500123, No. 531205006, and No. 533305009. X.D. is also supported by the Research Council of Finland, the Centre of Excellence in Quark Matter (Project No. 364191).
We also acknowledge the support of the High Performance Computing Center, Tsinghua University.

\section*{Data Availability}
The data that support the findings of this article are openly
available~\cite{zheng_2026_19842486}.

\appendix
\section{Triadiagonal Matrix Algorithms}
\label{sec:app:tma}
In this work, we solve the discrete Schr\"odinger equation according to Eq.~\eqref{eq:discrete_Sch}, which contains the unknown wave function ($\{u_j^{n+1}\}_j$), corresponding to the new time step, on both sides. In this appendix section, we discuss how we solve it efficiently using the TMA (also known as the Thomas algorithm).

Solving Eq.~\eqref{eq:discrete_Sch} can be regarded as solving the linear equation $\mathbf{u} \equiv \{u_j^{n+1}\}_j$ from $\mathbf{A}\cdot\mathbf{u} = \mathbf{v}$. $\mathbf{A}$ is an $N \times N$ dimensional tridiagonal matrix with the subdiagonal, diagonal, and superdiagonal (denoted as $a$, $b$, and $c$, respectively) as the only non-vanishing elements:
\begin{align}
     \mathbf{A} = \begin{bmatrix} 
b_1 & c_1 & 0 & 0 & \cdots & 0 \\
a_2 & b_2 & c_2 & 0 & \cdots & 0 \\
0 & a_3 & b_3 & c_3 & \cdots & 0 \\
\vdots & \vdots & \ddots & \ddots & \ddots & \vdots \\
0 & \cdots & 0 & a_{N-1} & b_{N-1} & c_{N-1} \\
0 & \cdots & 0 & 0 & a_N & b_N \\
\end{bmatrix}\label{eq:Amat}.
\end{align}
Here $a_j = c_j = -1$ for all $j$'s,
\begin{align}
b_j =\,& 
    \frac{2 m_b \Delta r ^2}{i \Delta t} + 2 + m_b \Delta r ^2 U_j^{n+1}.
\end{align}
Meanwhile, $\mathbf{v}$ is an $N$-dimensional vector determined by the (known) wave function of the last time step, whose $j$th element is given by
\begin{align}
v_j =\,& 
    \Big(\frac{2 m_Q \Delta r ^2}{i \Delta t} - 2 - m_b \Delta r ^2 U_j^{n}\Big) u_{j}^{n} + u_{j-1}^{n} + u_{j+1}^{n}.
\end{align}

The basic idea of TMA is to convert the tridiagonal matrix into the product of a lower triangular matrix and an upper triangular matrix, $\mathbf{A} = \mathbf{L}\cdot\mathbf{U}$, and then solve the linear equation system through back substitution, i.e., find $\mathbf{v}'$ so that $\mathbf{L}\cdot \mathbf{v}'=\mathbf{v}$, and solve $\mathbf{u}$ from $\mathbf{U}\cdot\mathbf{u}=\mathbf{v}'$. The lower-upper decomposition should take a simple form,
\begin{align}
\begin{split}
\mathbf{A} =\,& \mathbf{L}\cdot\mathbf{U} =
\begin{bmatrix} 
\alpha_1 & &  & \\
a_2 & \alpha_2 & &\\
& \ddots & \ddots & &\\
& &a_N & \alpha_N \\ 
\end{bmatrix}\cdot
\begin{bmatrix} 
1 & c'_{1} & &\\
&\ddots & \ddots & \\
& & 1 &  c'_{N-1} \\
& & & 1 \\
\end{bmatrix}
\end{split}\label{eq:Amat_LU}.
\end{align}
Comparison between Eqs.~\eqref{eq:Amat} and~\eqref{eq:Amat_LU} yields $\alpha_1=b_1$ $c'_{1}= c_{1}/b_{1}$ and 
\begin{align} 
\alpha_i = b_{i} - a_{i} \; c'_{i-1}\,,\qquad
c'_{i}= c_{i}/\alpha_i,
\end{align}
for $i = 2, 3, \ldots, N$. Then, $\mathbf{v}'\equiv \mathbf{L}^{-1}\cdot\mathbf{v}$ can be solved by
\begin{align} 
v'_{1} = v_{1}/ \alpha_{1}, \quad  v'_{i\geq2} = (v_{i} - a_{i} \; v'_{i-1})/\alpha_i,
\end{align}
and the solution $\mathbf{u}=\mathbf{U}^{-1}\cdot\mathbf{v}'$ is given by
\begin{align} 
    u_N = v'_{N},  \qquad
    u_{i<N} = v'_{i} - c'_{i} \; u_{i+1}.
\end{align}

The TMA has the advantage of having a computational complexity of $O(n)$, making it highly efficient for solving large-scale tridiagonal linear equation systems.

\section{Model Emulator}\label{sec:app:emulator}
Although the TMA has been invoked in solving the time-dependent Schr\"odinger equation, computing the $R_{AA}$'s given a potential (with a certain parameter set) remains time-consuming. Thus, we further deploy surrogate models to approximate this process. We first perform principal component analysis to find the combinations of observables that are most sensitive to changes in parameters and then employ a Gaussian process to describe such principal components. We provide details of both methods in this appendix section.

\subsection{Principal component analysis} \label{SectionPCA}
Using the Schr\"odinger equation, we have already obtained the database that includes both the parameter $\{ \boldsymbol{x}\}$ and its corresponding theoretical calculation $\{ \boldsymbol{y}(\boldsymbol{x}) \}$. 
A PCA starts with constructing the observable covariance matrix concerning the changes in model parameters,
\begin{align}
\mathbf{C}^{OO}_{ij} &\equiv \langle \delta y_i(\boldsymbol{x}) \delta y_j(\boldsymbol{x}) \rangle_{\boldsymbol{x}} 
 = \frac{1}{n} \sum_{m=1}^{n} \delta y_{i,m} \delta y_{j,m},
\end{align}
where $n$ represents the number of databases (e.g., $n=200$), and
\begin{align}
\delta y_{i,m} & \equiv y_{i}(\boldsymbol{x}_m) - \overline{y_{i}}, \qquad
\overline{y_{i}} \equiv \frac{1}{n} \sum_{m=1}^{n} y_{i}(\boldsymbol{x}_m).
\end{align}

PCA aims to find the principal components, $\mathrm{P}$, as linear combinations of the observables,
\begin{align}
P_i &= \sum_j U_{ij} \,\delta y_j,
\end{align}
so that the covariance matrix
\begin{align}
\mathbf{C}^{PP} &\equiv \langle P_i(\boldsymbol{x}) P_j(\boldsymbol{x}) \rangle_{\boldsymbol{x}} = \mathbf{U} \cdot \mathbf{C}^{OO} \cdot \mathbf{U}^{\dagger},
\end{align}
becomes diagonal. In other words, different $P_i$'s are not correlated with each other. As will be discussed in the next subsection, we use the Gaussian process to provide an interpolation-based approximation of $P_i(\boldsymbol{x})$ associated with uncertainty $\sigma_{P,i}(\boldsymbol{x})$. The prediction of the model can be further obtained as $\boldsymbol{y}(\boldsymbol{x}) = \mathbf{U}^{\dagger} \cdot \mathrm{P}(\boldsymbol{x}) + \overline{\boldsymbol{y}}$.

The likelihood distribution~\eqref{eq:likelihood} characterizing the probability given the difference between the experimental result and the prediction of the model $\Delta \boldsymbol{y}_{\boldsymbol{x}} \equiv \boldsymbol{y}(\boldsymbol{x}) - \boldsymbol{y}_{\mathrm{exp}}$ reads
\begin{align*}
\mathcal{P}(\boldsymbol{y}_{\mathbf{exp}}|\boldsymbol{y}(\boldsymbol{x}))   
= \frac{\exp\left[-\frac{1}{2} (\Delta \boldsymbol{y}_{\boldsymbol{x}})^{T} \cdot\Sigma^{-1}(\boldsymbol{x}) \cdot \Delta \boldsymbol{y}_{\boldsymbol{x}}\right]}{\sqrt{(2\pi)^{d} \det\left[\Sigma(\boldsymbol{x}) \right]}}.
\end{align*}
The total covariance matrix $\Sigma(\boldsymbol{x})$ is given by the summation of the experimental uncertainty ($\Sigma_{\mathrm{exp}}$) and the theoretical one,
\begin{align}
\Sigma(\boldsymbol{x}) &= \Sigma_{\mathrm{th}}(\boldsymbol{x}) + \Sigma_{\mathrm{exp}}.
\end{align}
While the latter is given by measurements, the former is calculated by incorporating the uncertainty of the Gaussian process $\Sigma_{\mathrm{Gauss}}$, where a factor of 2 is taken into account to represent the uncertainty from the Schr\"odinger evolution framework, i.e., $\Sigma_{\mathrm{th}}(\boldsymbol{x}) =2   \Sigma_{\mathrm{Gauss}}(\boldsymbol{x})$, with
\begin{align}
\Sigma_{\mathrm{Gauss}}(\boldsymbol{x}) &= \mathbf{U}^{\dagger} \cdot \mathrm{diag}\big(\sigma_{P,1}^2(\boldsymbol{x}),\sigma_{P,2}^2(\boldsymbol{x}), \cdots\big) \cdot \mathbf{U}.
\end{align}

\subsection{Gaussian processes}\label{chapterGP}
A GP provides probabilistic descriptions of functions in which both inputs and outputs are defined in a continuous domain. We demonstrate GP using the example of our estimation of a particular principal component in the preceding subsection $P(\boldsymbol{x})$, given that we have computed its values at discrete points $\{\boldsymbol{x}_1, \boldsymbol{x}_2, \ldots, \boldsymbol{x}_n\}$, with the values denoted as $\{P_1, \ldots P_n\}$, respectively.

In GP, one always needs to choose a kernel function, $\kappa(\boldsymbol{x}, \boldsymbol{x}')$, to characterize the two point correlation. An auxiliary kernel matrix is defined as
\begin{align}
\mathbf{K}\equiv\begin{bmatrix}
\kappa(\boldsymbol{x}_1,\boldsymbol{x}_1)&\cdots&\kappa(\boldsymbol{x}_1,\boldsymbol{x}_n)\\
\vdots&\ddots&\vdots&\\
\kappa(\boldsymbol{x}_n,\boldsymbol{x}_1)&\cdots&\kappa(\boldsymbol{x}_n,\boldsymbol{x}_n) 
\end{bmatrix} + \sigma^2_{\mathrm{wn}} \mathbf{I}_{n\times n},
\end{align}
where $\sigma^2_{\mathrm{wn}}$ is the white noise variance added to each point of the function.
For an arbitrary input, the value of the function of interest follows a normal distribution,
\begin{align}
    P(\boldsymbol{x}) \sim \mathcal{N}\big(\overline{P}(\boldsymbol{x}), \sigma^2(\boldsymbol{x})\big),
\end{align}
in which the mean value and the variance are respectively given by
\begin{align}
\overline{P}(\boldsymbol{x}) =\,& 
    \sum_{i,j=1}^{n} \kappa(\boldsymbol{x}, \boldsymbol{x}_i) (\mathbf{K}^{-1})_{ij} P_j\,,\\
\sigma^2(\boldsymbol{x}) =\,&
    \kappa(\boldsymbol{x},\boldsymbol{x}) - \sum_{i,j=1}^{n} \kappa(\boldsymbol{x},\boldsymbol{x}_i)(\mathbf{K}^{-1})_{ij} \kappa(\boldsymbol{x}_j,\boldsymbol{x})\,.
\end{align}
Obviously, at a precalculated parameter set, $\boldsymbol{x}_k$, one may easily find $\overline{P}(\boldsymbol{x}=\boldsymbol{x}_k) = P_k$ and $\sigma^2(\boldsymbol{x}=\boldsymbol{x}_k) = \kappa(\boldsymbol{x}_k,\boldsymbol{x}_k) - \mathbf{K}_{k,k} = 0$ when the white noise vanishes ($\sigma_{\mathrm{wn}}=0$).
In this work, we have chosen the standard radial basis function kernel function,
\begin{align}
\kappa(\boldsymbol{x},\boldsymbol{x}')=C^2\exp\Big(-\frac{\left\|\boldsymbol{x}-\boldsymbol{x}'\right\|_2^2}{2l^2}\Big),
\end{align}
where $\left\|\boldsymbol{x}-\boldsymbol{x}'\right\|_2 \equiv \sqrt{\sum_k(x_{k}-x_{k}')^2}$ is the norm to describe the spatial distance between the vectors $\boldsymbol{x}$ and $\boldsymbol{x}'$.
The autocorrelation coefficient $C$, the length scale $l$, as well as the white noise scale $\sigma_{\mathrm{wn}}$ are hyperparameters that should be optimized by using Markov-chain Monte Carlo sampling and maximizing the logarithmic marginal likelihood.

\section{Parameter Sampling}\label{sec:app:mcmc}
The implementation of PCA and GP, introduced in the preceding section, has allowed for a highly efficient computation of the posterior density for a given parameter set, $\mathcal{P}_{\mathrm{posterior}}(\boldsymbol{x})$. Nevertheless, to obtain the full $\mathcal{P}_{\mathrm{posterior}}(\boldsymbol{x})$, an efficient algorithm is still needed to explore the parameter space. The Markov-chain Monte Carlo method is widely used for this purpose, sampling the parameter points according to $\mathcal{P}_{\mathrm{posterior}}(\boldsymbol{x})$ and then estimating the most essential properties of the distribution. 
In what follows we will introduce two commonly adopted methods in MCMC---the Metropolis-Hasting method to review the basic idea of parameter sampling and the Hamiltonian Monte Carlo for an efficient method.

\subsection{Metropolis--Hastings method}\label{Hastings}
As one of the most important methods to realize the Markov-chain Monte Carlo sampling, the Metropolis--Hastings method enables us to derive a stationary distribution of the parameter (such as the posterior) by utilizing the rejection sampling algorithm. Before we pay attention to the algorithm itself, the concept of Monte Carlo sampling should be introduced first. Next, a few basic concepts arising from the Markov chain will be demonstrated, followed by the derivation of the Hastings method.

In practice, the multidimensional space of the parameter (denoted as $\Omega_{\boldsymbol{x}}$) is much larger than the credible interval (referred to as $\Omega^*$) that is relevant to the Bayesian estimation. Consequently, an efficient exploration should be conducted around the target interval $\Omega^*$, but it's of no necessity to scan the entire parameter space $\Omega_{\boldsymbol{x}}$.

Reconstruction of the posterior likelihood distribution could be performed by constructing an independent and identically distributed random sampling sequence $\{\boldsymbol{x}_{(i)}\}$. That is, $\mathcal{P}_{\mathrm{posterior}}(\boldsymbol{x}_{(i)})\mathrm{d}^d\boldsymbol{x}_{(i)}$ is constant and independent of $i$, with $\mathrm{d}^d\boldsymbol{x}_{(i)}$ being the volume element corresponding to the sample $\boldsymbol{x}_{(i)}$. 

To utilize sampling methods well developed in statistical physics, let us make the analogy that the parameter distribution corresponds to a ``virtual'' physical system, in which each parameter set ($\boldsymbol{x}$) corresponds to a physical state, and the thermal equilibrium distribution of the system corresponds exactly to our target posterior likelihood distribution, $\mathcal{P}_{\mathrm{posterior}}(\boldsymbol{x})$. Thus, obtaining the parameter chain, $\{\boldsymbol{x}_{(i)}\}$, would be equivalent to generating a thermal ensemble of states.  

The ensemble can be generated utilizing the method of Markov chains, in which states are evolved according to transitions that eventually drive the system towards the desired thermal equilibrium. The Markov-chain methods make the no-memory assumption, in which the new state depends only on the one in the last step and not on the earlier ones in the chain, i.e.,
\begin{align}
\mathcal{P}(\boldsymbol{x}_{(t+1)}|\{\boldsymbol{x}_{(1)},  \ldots, \boldsymbol{x}_{(t)}\}) = \mathcal{P}(\boldsymbol{x}_{(t+1)}|\boldsymbol{x}_{(t)}).
\end{align}

We denote the distribution of states at the $t$th step as $\mathcal{P}_t(\boldsymbol{x}) \equiv \mathcal{P}(\boldsymbol{x}_{(t)}=\boldsymbol{x})$, and the transition rate from a state ($\boldsymbol{x}'$) in the last step to another one ($\boldsymbol{x}$) in the new step,
\begin{align}
\mathcal{T}_{\boldsymbol{x}' \to \boldsymbol{x}} \equiv \mathcal{P}(\boldsymbol{x}_{(t+1)} = \boldsymbol{x}|\boldsymbol{x}_{(t)} = \boldsymbol{x}').
\end{align}
Obviously, the distribution of states evolves according to
\begin{align}
    \mathcal{P}_{t+1}(\boldsymbol{x})
    = \int \mathrm{d}^d \boldsymbol{x}'\;
    \mathcal{P}_{t}(\boldsymbol{x}')
    \mathcal{T}_{\boldsymbol{x}' \to \boldsymbol{x}}\,.
\end{align}
They are expected to converge to the desired thermal distribution, 
\begin{align}
    \lim_{t\to\infty} \mathcal{P}_{t+1}(\boldsymbol{x}) = \lim_{t\to\infty} \mathcal{P}_{t}(\boldsymbol{x}) = \mathcal{P}_{\mathrm{posterior}}(\boldsymbol{x})\,,
\end{align}
which automatically leads to the detailed balance
\begin{align}
    \mathcal{P}_{\mathrm{posterior}}(\boldsymbol{x}) \mathcal{T}_{\boldsymbol{x} \to \boldsymbol{x}'}
    =\mathcal{P}_{\mathrm{posterior}}(\boldsymbol{x}') \mathcal{T}_{\boldsymbol{x}' \to \boldsymbol{x}}\,,
\end{align}
and can be satisfied if the transition rates are determined according to
\begin{align}
    \mathcal{T}_{\boldsymbol{x} \to \boldsymbol{x}'} = \mathrm{min}\Big(1, \frac{\mathcal{P}_{\mathrm{posterior}}(\boldsymbol{x}')}{\mathcal{P}_{\mathrm{posterior}}(\boldsymbol{x})}\Big)\,.
    \label{eq:acceptance_probability}
\end{align}

Thus, the Metropolis--Hastings method samples an ensemble of states, i.e., a chain of parameters, 
$\{\boldsymbol{x}_{(1)},\boldsymbol{x}_{(2)}, \ldots \}$, by repeating the following iterations,
\begin{itemize}
\item[1.] 
    Start from the very last point of the list, denoted $\boldsymbol{x}$, and sample a new set of parameters $\boldsymbol{x}'$ by taking a random walk around it.
\item[2.]
    Sample a random number $\xi\in[0,1]$, accept the new parameter (append it to the $\{\boldsymbol{x}_{(1)},\boldsymbol{x}_{(2)}, \ldots \}$ list) if
\begin{align*}
    \xi < \frac{\mathcal{P}_{\mathrm{posterior}}(\boldsymbol{x}')}{\mathcal{P}_{\mathrm{posterior}}(\boldsymbol{x})}.
\end{align*}
\end{itemize}
With sufficient parameter sets collected, their distributions converge to the posterior distribution. Nevertheless, if one takes too large a step in the random walk, $\boldsymbol{x}'$ would be far from the credible interval, and the acceptance rate would be very small since $\mathcal{P}_{\mathrm{posterior}}(\boldsymbol{x}') \approx 0$; conversely, if the step size is too small, then there exist strong correlations between different $\boldsymbol{x}_{(t)}$'s in the chain. In practice, one usually takes a medium step size to ensure a reasonable acceptance rate while maintaining controllable autocorrelation. The latter is further eliminated by only keeping a small portion of the generated chain -- a minimal interval is required for the samples that are adopted.

Obviously, the method of Metropolis--Hastings sampling is inefficient. Meanwhile, the parameter update strongly depends on the initial position of the parameter set, and it may encounter the problem of local minima. The Hamiltonian Monte Carlo sampling method is capable of covering the multimodal distributions and exhibits high numerical efficiency.

\subsection{Hamiltonian Monte Carlo}\label{HamiltonianMonteCarlo}
We continue to use the language of thermodynamics and statistical physics for a consistent discussion. Although the HMC sampling method shares the same form of acceptance probability~\eqref{eq:acceptance_probability} with Metropolis--Hastings, it proposes the next state based on the previous one using a physics-motivated strategy.

The HMC method defines an effective potential based on the desired thermal distribution,
\begin{align}
    V(\boldsymbol{x}) = -T\,\ln \mathcal{P}_{\mathrm{posterior}}(\boldsymbol{x})\,,
\end{align}
introduces the grand momentum ($\boldsymbol{p}$) that conjugates to the grand coordinate ($\boldsymbol{x}$) and considers a system governed by the Hamiltonian
\begin{align}
H(\boldsymbol{x},\boldsymbol{p}) &= \frac{\boldsymbol{p}\cdot\boldsymbol{p}}{2m} + V(\boldsymbol{x}).
\end{align}
When such a system thermalizes, the phase-space distribution is given by
\begin{align}
    \mathcal{P}_{\mathrm{eq}}(\boldsymbol{x}, \boldsymbol{p}) = \frac{e^{-H(\boldsymbol{x},\boldsymbol{p})/T}}{Z}\,.
    \label{eq:HMC_equilibrium}
\end{align}
Noting that the distribution can be separated into the product of a momentum-dependent sector and a coordinate-dependent sector, one may show that the marginal distribution of the coordinates recovers the desired posterior likelihood distribution of the model parameters,
\begin{align}
    \int \mathcal{P}_{\mathrm{eq}}(\boldsymbol{x}, \boldsymbol{p}) \mathrm{d}^d \boldsymbol{p} = \mathcal{P}_{\mathrm{posterior}}(\boldsymbol{x})\,.
\end{align}
Thus, in the accumulative sampling of the Markov chain, the HMC evolves not only $\boldsymbol{x}$ but also $\boldsymbol{p}$, so that eventually an ensemble can be collected that follows the thermal equilibrium distribution~\eqref{eq:HMC_equilibrium}.

An HMC iteration starts from the newest coordinate point in the chain, denoted as $\boldsymbol{x}_{(t)}$, and randomly samples the momentum from the thermal distribution, $\boldsymbol{p} \sim (2\pi m T)^{-\frac{d}{2}}\,e^{-\frac{\boldsymbol{p}\cdot \boldsymbol{p}}{2mT}}$.
With the phase-space point $(\boldsymbol{x} = \boldsymbol{x}_{(t)}, \boldsymbol{p})$ being the initial condition, the HMC evolves the phase-space trajectory according to Hamilton's canonical equations
\begin{align}
\dot{x}_i = \frac{\partial H}{\partial p_i} = \frac{p_i}{m}, 
\qquad \dot{p}_i = -\frac{\partial H}{\partial x_i} = - \frac{\partial V(\boldsymbol{x})}{\partial x_i}, \label{eq:canonical_equation}
\end{align}
for a certain time interval ($t_{\Delta}$), it arrives at the final state labeled $(\boldsymbol{x}', \boldsymbol{p}')$. Obviously, the phase-space distributions at the $t$th and $(t+1)$th steps satisfy
\begin{align}
    \mathcal{P}_{t+1}(\boldsymbol{x}', \boldsymbol{p}') = 
    \mathcal{P}_{t}(\boldsymbol{x}, \boldsymbol{p}) J(\boldsymbol{x}', \boldsymbol{p}')\,,
\end{align}
where $\boldsymbol{x}$ and $\boldsymbol{p}$ should be treated as the (inverse) functions of $(\boldsymbol{x}', \boldsymbol{p}')$, i.e., $\boldsymbol{x} = \boldsymbol{x}(\boldsymbol{x}', \boldsymbol{p}')$ and $\boldsymbol{p} = \boldsymbol{p}(\boldsymbol{x}', \boldsymbol{p}')$, with $J(\boldsymbol{x}', \boldsymbol{p}')$ being the corresponding Jacobian,
\begin{align}
    J(\boldsymbol{x}', \boldsymbol{p}') \equiv
    \mathrm{det}\begin{bmatrix}
    \vspace{2mm}
        \frac{\partial x_i}{\partial x'_{j}} & \frac{\partial x_i}{\partial p'_{j}} \\
        \frac{\partial p_i}{\partial x'_{j}} & \frac{\partial p_i}{\partial p'_{j}}
    \end{bmatrix}\,.
\end{align}

In the idealized case assuming infinite numerical precision, Eq.~\eqref{eq:canonical_equation} conserves the energy and all points on the trajectory have the same Hamiltonian. In particular, 
\begin{align}
    H(\boldsymbol{x}', \boldsymbol{p}') = H(\boldsymbol{x}, \boldsymbol{p})\,.\label{eq:energy_conservation}
\end{align}
Meanwhile, from the Liouville theorem, one can show that the element volume is conserved, i.e., the Jacobian of the transformation remains unity, $J(\boldsymbol{x}', \boldsymbol{p}')=1$. Therefore, once we have approached the desired distribution in the $t$th step, $\mathcal{P}_{t}(\boldsymbol{x}, \boldsymbol{p}) = \mathcal{P}_{\mathrm{eq}}(\boldsymbol{x}, \boldsymbol{p})$, one can show that the following phase-space points also follow the thermal distribution,
\begin{align}
    \mathcal{P}_{t+1}(\boldsymbol{x}', \boldsymbol{p}')
    = \mathcal{P}_{t}(\boldsymbol{x}, \boldsymbol{p})
    = \frac{e^{-\frac{H(\boldsymbol{x},\boldsymbol{p})}{T}}}{Z} 
    = \frac{e^{-\frac{H(\boldsymbol{x}',\boldsymbol{p}')}{T}}}{Z}\,.
\end{align}

In practice, Hamilton's canonical equation~\eqref{eq:canonical_equation} needs to be solved numerically with discrete time steps, and the gradients of the effective potential might not be analytically known. In this work,  the functions for computing numerical gradients in \textsc{scipy.optimize} are used to handle the partial derivative terms. The conservation of the volume of phase space is ensured when the Leap Frog scheme is utilized to solve Hamilton's canonical equations, which can be expressed in the following formulas:
\begin{align}
\begin{split}
\hat{p}_i\big(t+\frac{\delta_t}{2}\big) &=  \hat{p}_i(t) - \frac{\delta_t}{2} \frac{\partial V}{\partial \hat{x}_i}(t),\\
\hat{x}_i(t+\delta_t)&=\hat{x}_i(t)+\delta_t\frac{\hat{p}_i\big(t+\frac{\delta_t}{2}\big)}m, \\
\hat{p}_i(t+\delta_t) &=  \hat{p}_i(t+\frac{\delta_t}{2}) - \frac{\delta_t}{2} \frac{\partial V}{\partial \hat{x}_i}(t+ \delta_t).
\end{split}
\label{eq:Leap_Frog}
\end{align}
By iteratively solving the four equations above, we can update the parameters. Specifically, we take $\boldsymbol{x}'= \hat{\boldsymbol{x}}(t+N_t\delta_t)$ and $\boldsymbol{p}'= \hat{\boldsymbol{p}}(t+N_t\delta_t)$, where $t_{\Delta} \equiv N_t\delta_t$ is the time interval mentioned above.

Considering the finite numerical accuracy, energy conservation~\eqref{eq:energy_conservation} does not necessarily hold. This inaccuracy can be corrected by further introducing an additional acceptance rate. All in all, the practical HMC method samples the parameter chain,
$\{\boldsymbol{x}_{(1)},\boldsymbol{x}_{(2)}, \cdots \}$, by repeating the following iterations,
\begin{itemize}
\item[1.] 
    Start from the very last point of the list, denoted $\boldsymbol{x}$, and sample the momentum according to the thermal distribution,
    $p_i \sim \mathcal{N}(p_i;\mu=0,\sigma^2=mT)$ for $i=1,\cdots,d$.
\item[2.]
    Use the Leap Frog method~\eqref{eq:Leap_Frog} and solve the phase-space evolution starting from $(\boldsymbol{x}, \boldsymbol{p})$. After $N_t$ time steps, we arrive at another point in the phase space, denoted as $(\boldsymbol{x}', \boldsymbol{p}')$.
\item[3.]
    Sample a random number $\xi\in[0,1]$; accept $\boldsymbol{x}'$ (append it to the $\{\boldsymbol{x}_{(1)},\boldsymbol{x}_{(2)}, \cdots \}$ list) if
\begin{align*}
    \xi < \exp\left[-\frac{H(\boldsymbol{x}',\boldsymbol{p}')-H(\boldsymbol{x},\boldsymbol{p})}{T}\right].
\end{align*}
\end{itemize}
In the above, the effective temperature ($T$) and the mass ($m$) are parameters setting the scale of thermal momentum and the relative update step between the grand coordinates and grand momenta.

\begin{figure}[!b]
    \centering
    \includegraphics[width=.45\textwidth]{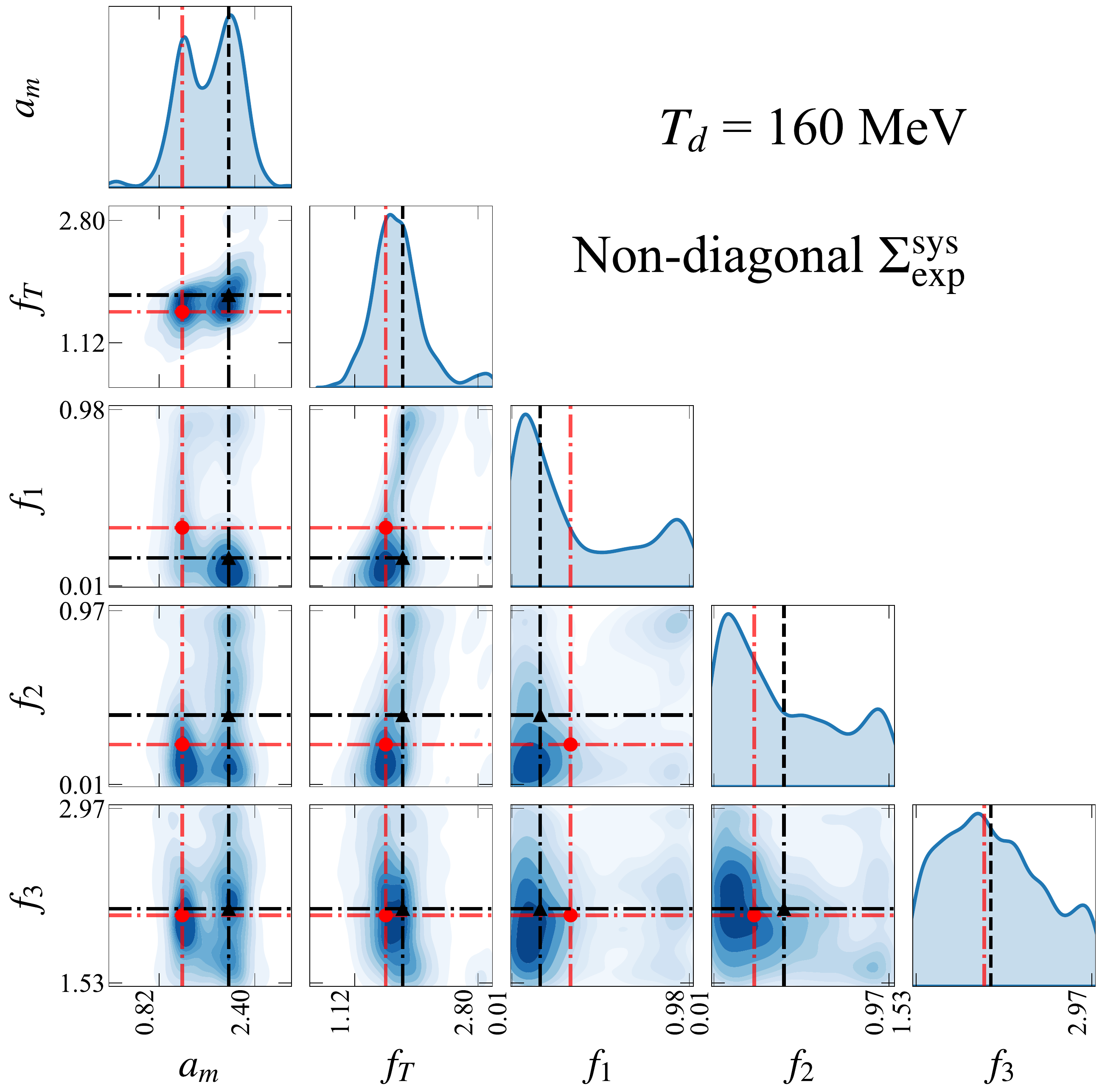}
    \caption{Same as Fig.~\ref{fig:Posterior_160}, but for a nondiagonal experimental systematic covariance matrix.}
    \label{fig:Posterior_160_nondiag}
\end{figure}

\begin{figure}[!t]
    \centering
        \includegraphics[width=0.45\textwidth]{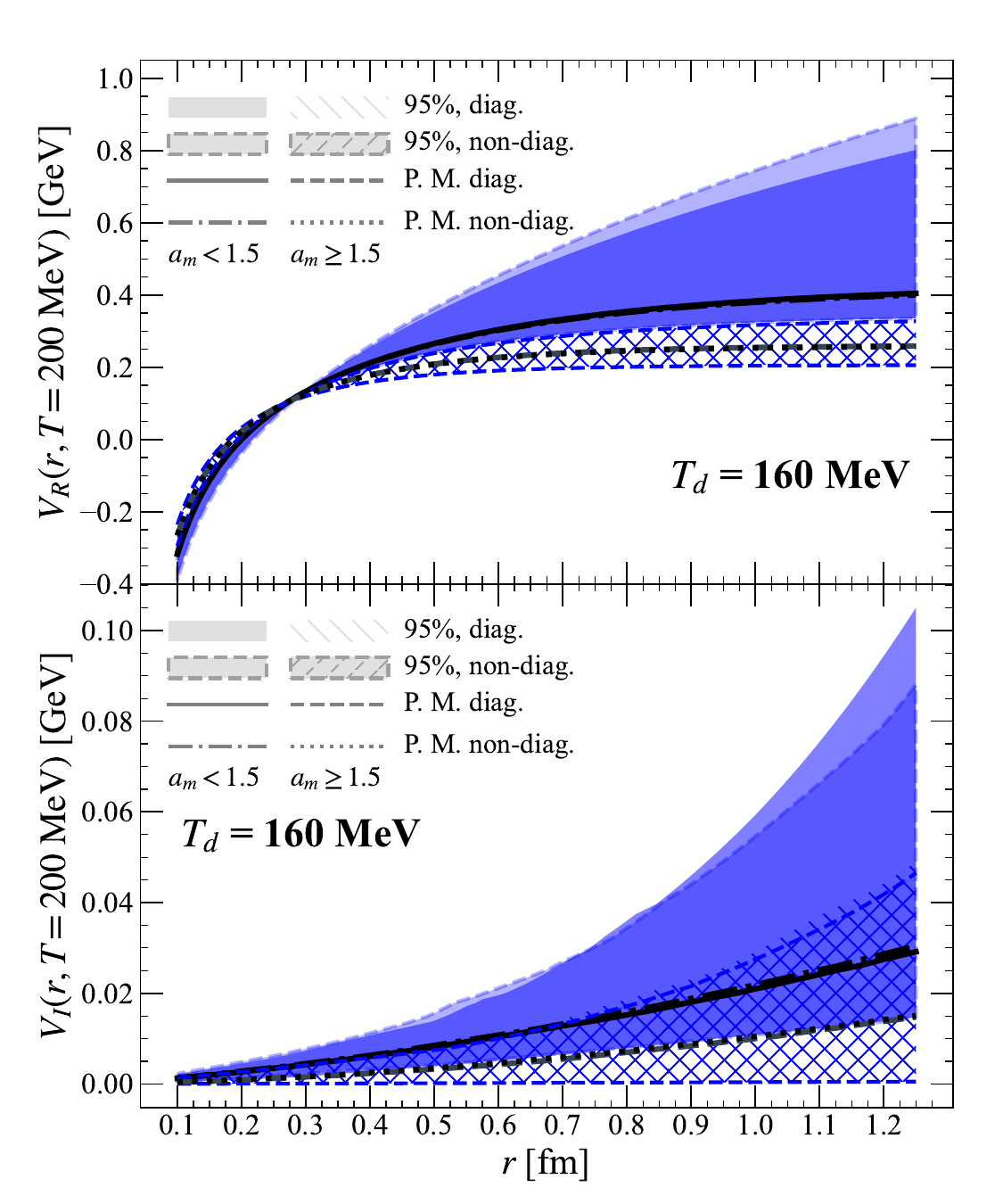}
    \caption{Comparison of the 95\% confidence intervals obtained with diagonal systematic uncertainties (solid blue band) and nondiagonal systematic uncertainties (dashed blue band). The plot details and drawing style are consistent with those of \protect{Fig.~\ref{fig:EffectvePotential_160}.}
    \label{fig:EffectvePotential_nondiag}}
\end{figure}

\section{Nondiagonal Experimental Systematic Covariance Matrix}\label{Sigma_syst}
Current LHC experiments report both correlated and uncorrelated systematic uncertainties.
While we keep only diagonal systematic uncertainties in the Bayesian analysis reported in the main text, here we study the effect of the nondiagonal elements.

For measurements of the same physical observable from a single experimental collaboration, the experiment usually introduces global uncertainty between different data points through, e.g., the pp luminosity.
Consequently, we can represent the systematic uncertainties in a block-diagonal form.
The full systematic uncertainties can be decomposed into two components:
The first is the diagonal terms provided by the uncorrelated systematic errors, and the second corresponds to the correlations cause by the global scaling, i.e.,
\begin{align}
\Sigma_{\mathrm{exp}}^{\mathrm{sys}} &= \Sigma_{\mathrm{diag}}^{\mathrm{sys}} + \Sigma_{\mathrm{corr}}^{\mathrm{sys}}, \\
\Sigma_{\mathrm{diag}}^{\mathrm{sys}} &= \mathrm{diag} \left(\delta_{\mathrm{sys},1}^2, \delta_{\mathrm{sys},2}^2, \cdots, \delta_{\mathrm{sys},d}^2 \right),
\end{align}

where $\delta_{\mathrm{sys},i},~i=1,2,\ldots,d$ are the uncorrelated uncertainties, and $\Sigma_{\mathrm{corr}}^{\mathrm{sys}}$ is the block-diagonalized matrix representing the correlations between different dimensions of the same physical observable. Specifically, we may present the explicit form of the matrix elements of $\Sigma_{\mathrm{corr}}^{\mathrm{sys}}$, i.e.,
\begin{equation}
\Sigma_{\mathrm{corr},i,j}^{\mathrm{sys}} =\sigma_\lambda^2 ~y_{\mathrm{exp},i} ~y_{\mathrm{exp},j}.
\end{equation}
Here $\sigma_{\lambda}$ represents the global scaling uncertainty of a specific physical observable, and the subscript $\lambda$ is used to distinguish different physical measurements from different experimental collaborations. $y_{\mathrm{exp},i}$ denotes the experimental value of the physical observable associated with matrix index $i$.

\begin{figure}[!h]
    \centering
        \includegraphics[width=0.45\textwidth]{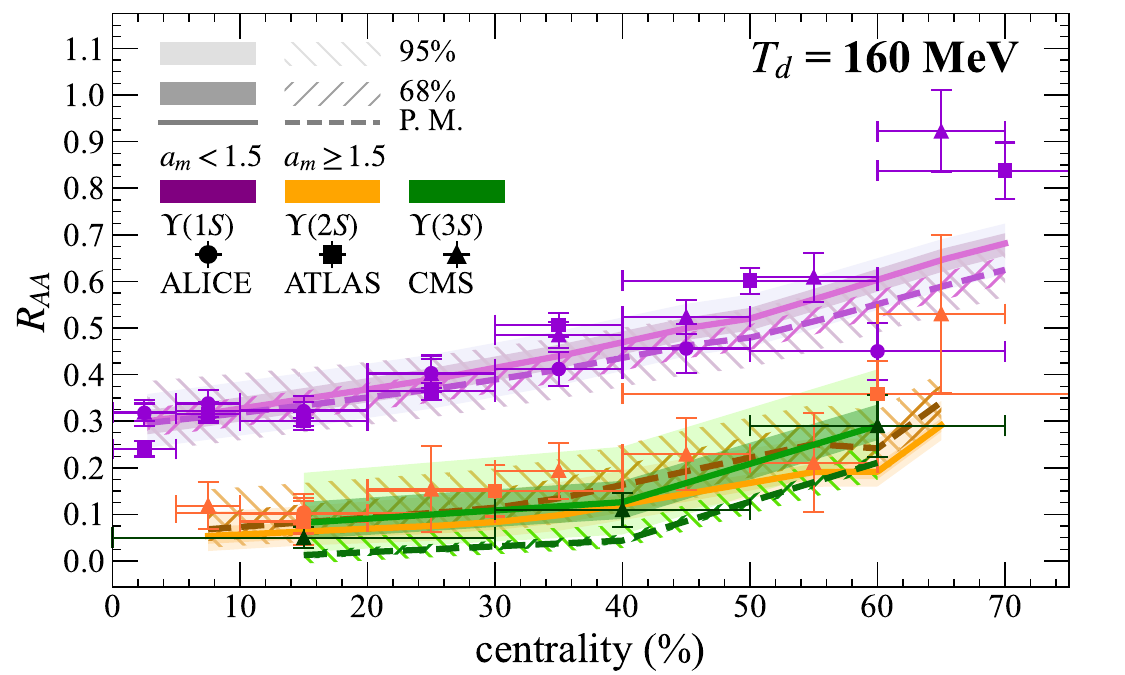}
        \includegraphics[width=0.45\textwidth]{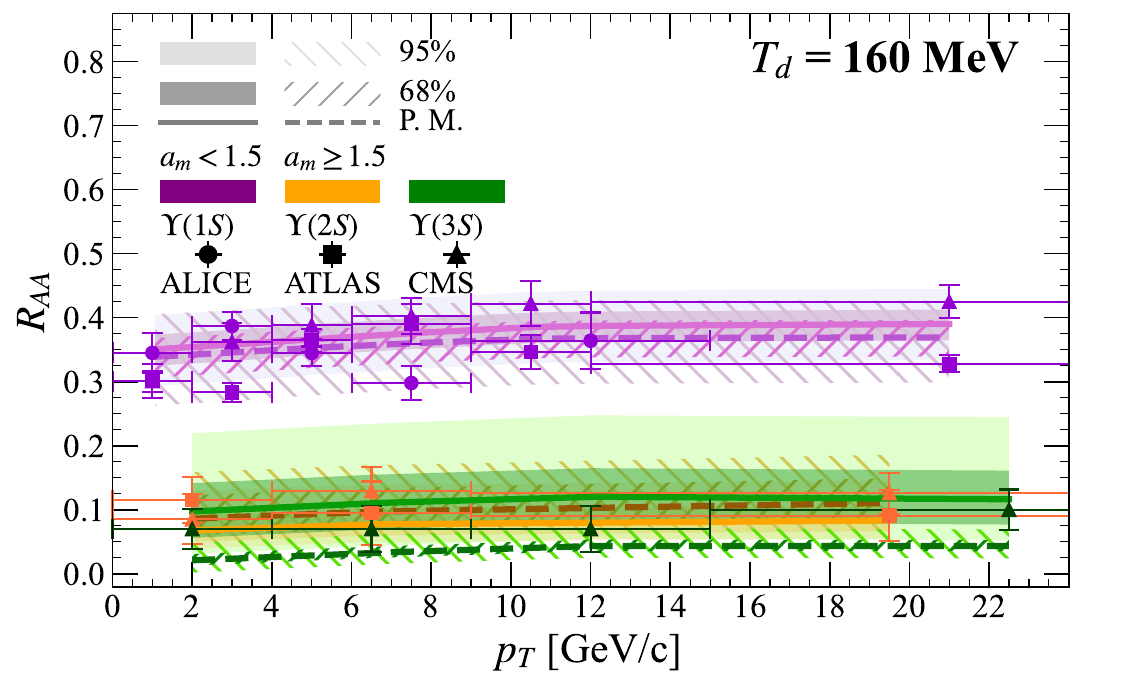}
    \caption{Same as \protect{Fig.~\ref{fig:RAA_160MeV}} but for nondiagonal systematic uncertainties $\Sigma_{\mathrm{exp}}^{\mathrm{sys}}$.
    \label{fig:RAA_nondiag}}
\end{figure}

Using this treatment for full systematic uncertainties, we have performed the Bayesian analysis at $T_d=160$ MeV, with the corresponding results shown in Figs.~\ref{fig:Posterior_160_nondiag},~\ref{fig:EffectvePotential_nondiag}, and~\ref{fig:RAA_nondiag}. We find that including nondiagonal systematic uncertainties in the present work does not significantly affect the results of the Bayesian analysis, especially when it comes to the more favored case with $a_m \geq 1.5$. For this reason, we adopt the diagonal matrix form for the systematic uncertainties throughout the main text.

\FloatBarrier
\bibliography{ref}

\end{document}